%% file: main.tex
\documentclass[11pt]{article}

\usepackage[margin=1.15in]{geometry}
\usepackage[T1]{fontenc}
\usepackage{microtype}
\usepackage{setspace}
\usepackage{booktabs}
\usepackage{longtable}
\usepackage{array}
\usepackage{graphicx}
\usepackage[font=it]{caption}
\usepackage[numbers,square]{natbib}
\usepackage{xurl}
\usepackage[hidelinks]{hyperref}
\usepackage{pgfkeys}

\newcommand{\setstat}[2]{\pgfkeys{/stat/#1/.initial={#2}}}
\newcommand{\stat}[1]{\pgfkeysifdefined{/stat/#1}{\pgfkeysvalueof{/stat/#1}}{\PackageError{paper}{Undefined statistic: #1}{Check paper_stats.tex.}}}
\input{paper_stats.tex}

\title{Reducing SNAP Payment Errors at Lower Cost With AI: \\
A Simulation Study}
\author{Alex Chohlas-Wood\\
\small Department of Applied Statistics, Social Science, and Humanities\\
\small New York University}
\date{}

\begin{document}
\maketitle

\begin{abstract}
\noindent
New federal legislation will require states with high payment error rates
in the Supplemental Nutrition Assistance Program (SNAP)
to pay a share of benefit costs,
potentially costing some states over a billion dollars per year.
These costs create substantial pressure to improve payment accuracy.
One strategy to improve payment accuracy is \textit{preauthorization review},
in which staff double-check applications and correct mistakes before benefits are issued.
But limited staff capacity makes it important to target these reviews effectively.
One leading targeting approach uses statistical risk models
to prioritize cases likely to contain errors.
Another novel approach uses AI to flag specific potential discrepancies for reviewers to check.
For states under pressure to reduce error rates,
it is unclear which approach---or combination of approaches---would reduce payment error rates most effectively.
We compare these strategies through simulations using public SNAP quality control data.
Under our main assumptions,
risk-based review alone costs about \stat{results.risk.combined.cost.ratio.median.words} times as much
as combining AI flags with risk scores.
This cost advantage persists across a wide range of plausible assumptions.
Our findings suggest that AI-assisted preauthorization review 
could be a key tool to help states
avoid substantial cost-sharing obligations.
\end{abstract}

\section*{Introduction}

The Supplemental Nutrition Assistance Program (SNAP)
helps low-income households in the United States purchase food.
Under this program, state agencies determine whether households are eligible
and how much assistance they should receive,
based on information about their income, expenses, and household circumstances.
But sometimes they make mistakes.
The federal government has historically paid the full cost of SNAP benefits,
giving it a financial stake in the accuracy of these determinations.
As a result, the federal government requires states
to conduct detailed quality control reviews of a random sample of cases,
and the U.S. Department of Agriculture uses these reviews
to estimate each state's annual \textit{payment error rate} (PER)~\citep{snap_qc_process}.
PER measures the total dollars overpaid or underpaid
as a percentage of total benefits issued.

The federal budget law H.R.~1, enacted in 2025,
sharply raised the financial stakes of these payment errors for states.
It followed a substantial increase in PER in states across the country
immediately after the pandemic;
despite recent modest declines, 
PER generally remains above pre-pandemic levels~\citep{snap_official_per}.
Under the new law, states will for the first time be required
to pay a portion of SNAP benefit costs
if their payment error rate is at or above \stat{per.cost.sharing.threshold}\%.
Their required contribution starts at
\stat{benefit.share.low}\% of their total SNAP benefit spending,
with higher error rates triggering larger contributions~\citep{snap_benefit_cost_sharing}.
As a result of this new requirement, starting in October 2027,
some states could face hundreds of millions or even billions of dollars
in new annual costs 
if they do not reduce their PER~\citep{california_snap_cost_sharing}.
Reducing PER below the threshold allows states to avoid these charges,
but it may be difficult for states to achieve in practice.
As an alternative,
some states have considered restricting eligibility or withdrawing from SNAP entirely
to avoid these new costs~\citep{aphsa_snap_survey}.
Helping states reduce payment errors could therefore protect
both state budgets and households' access to food assistance.

States are pursuing several ways to reduce payment errors,
including improving staff training,
clarifying instructions for applicants,
and modernizing eligibility systems~\citep{aphsa_snap_survey}.
One particularly promising approach is \textit{preauthorization review}:
after a caseworker makes a preliminary eligibility and benefit determination,
an experienced reviewer checks the case for errors
before benefits are issued.
Correcting errors at this stage prevents incorrect payments
that would otherwise contribute to the state's PER~\citep{snap_review_practices}.
But these checks require additional staff time,
and states lack the staff capacity to thoroughly double-check every detail of every case.
Agencies must therefore decide which cases warrant preauthorization review
and which specific details staff should check in the cases they review.

One reasonable strategy for preauthorization review
would be to prioritize review 
when it is likely that a case contains a payment error.
Statistical risk models and AI-based flagging offer two different approaches
to identifying such cases.
Risk models estimate which cases are most likely to contain errors
based on historical patterns of where such errors arise.
Many states are already using predictive models 
to target SNAP case reviews~\citep{aphsa_snap_survey}.
On the other hand, 
generative AI could review case materials 
and flag specific potential discrepancies for reviewers to check.
But generative AI is relatively new,
and its effectiveness in identifying SNAP payment errors remains largely unproven.
These approaches could also complement each other:
risk scores could help agencies decide which cases to review first,
while AI-based flags could help reviewers quickly locate errors within those cases.

States choosing among these approaches may be uncertain
about where to devote their time and resources,
and especially whether investing in AI would be worthwhile.
We investigate this question by simulating the potential impact 
of risk-based, AI-based, and combined preauthorization review.
Our aim is to understand 
the potential costs and benefits 
of these different approaches
under a range of plausible assumptions.
Our simulations suggest that combining AI flags with risk scores
could make preauthorization review substantially less costly
than relying on risk scores alone,
a pattern that holds across a wide range of assumptions.
These findings suggest that AI-assisted review has the potential
to help states substantially reduce payment error rates
and avoid costly cost-sharing obligations.

\section{Data}
\label{sec:data}

Our simulations are based on
the U.S. Department of Agriculture's
public-use SNAP Quality Control Data~\citep{snap_qc_data}.
This dataset is created through the following process.
Each month, state SNAP administrators draw a random sample
of households receiving SNAP benefits.
State auditors then conduct an intensive review of each selected case:
they re-examine case files,
re-interview participants,
and collect new information on household circumstances.
They use this information to determine whether the household
was actually eligible to receive benefits,
and whether they received the correct benefit amount.
Payments that differed from the correct benefit amount
are deemed a payment error
if they are above a tolerance threshold
(\$\stat{per.tolerance} in FY\stat{evaluation.year}).
Later,
USDA starts with this random sample of households,
filters out households found to be ineligible,
and compiles the remaining households into public-use files.
The public use files also include sampling weights
so downstream users can estimate characteristics
of each state's full eligible SNAP caseload.

Each record describes a household's demographics,
income, expenses, and benefits,
along with any payment errors identified during the review.
For each recorded error,
the data identify the ``element of variance''
(where the error occurred in the case)
and the ``primary cause of variance''
(how the error occurred).
For example, if a caseworker entered a household's wages incorrectly,
the element would be ``wages and salaries''
and the cause would be ``data entry and/or coding error.''
If the agency failed to account for housing expenses the household reported,
the element could be ``shelter deduction''
and the cause would be ``reported information disregarded or not applied.''
These attributes let us describe both
the type of payment error
and its cause.
For our simulations,
we group related elements of variance into ``categories''
that represent parts of a case a reviewer could check together.
For example, our ``earned income'' category includes errors
labeled with the elements ``wages and salaries,'' ``other earned income,''
and ``earned income deductions.''
Appendix~\ref{apx:review-categories} lists the full category mapping.

For our simulations,
we calculate state payment error rates (PERs)
directly from the public-use records,
even though the public-use records exclude ineligible households,
whose payments are included in the official PER.
As a check, we compare these estimates
with USDA's published state PERs~\citep{snap_official_per}.
Our estimates broadly track the official rates
but are generally
a few percentage points lower (Figure~\ref{fig:per-comparison}).
They therefore provide a rough proxy for official PER,
although the lower starting rates may make it easier
to reach the cost-sharing threshold in our simulations.
We also use USDA's annual state benefit-spending totals~\citep{snap_state_totals}
to translate our simulated PER
into cost-sharing penalties
if our simulated PER is at least \stat{per.cost.sharing.threshold}\%.

Our analysis draws on public-use records
from fiscal years \stat{analysis.years}
(FY\stat{evaluation.year} is the most recent year available~\citep{snap_qc_data}).
For simplicity,
we restrict our analysis to the \stat{sample.states.count} contiguous states
and the District of Columbia,
which share consistent maximum SNAP benefit amounts for each household size.
When simulating payment error correction,
we additionally exclude \stat{sample.excluded.households.percent} of households
included in the public use data
because they are
either subject to special benefit calculations
or have incomplete or inconsistent records.
The FY\stat{evaluation.year} households
that remain after these filters form our simulation sample.
For thoroughness, households excluded from our simulation sample
are reintroduced when calculating simulated error totals.

To simulate state use of risk-based review,
we adapt the publicly available SNAP risk model code
developed by Eric Giannella at the Better Government Lab (BGL),
which we call the ``BGL SNAP risk model''~\citep{snap_risk_code}.
The provided code trains statistical models
to identify households at higher risk of payment errors.
We retain its data preparation procedures and prediction model specification
and use the predicted probabilities to rank households for review.
However,
we train on a narrower set of years (FY\stat{training.years})
than the BGL SNAP risk model,
since we reserve FY\stat{evaluation.year} for evaluation.
We describe the model in more detail in the methods section.

Finally, published accounts from the Safety Net Response Network
provide rough benchmarks
for some of the assumptions we make in our analysis~\citep{snap_review_practices,snap_review_case_study}.
One case study in particular describes a state refining its selection rules.
Using historical data and estimates of which errors could be corrected,
the state projected
that conducting these reviews
under \stat{benchmark.targeting.rules.count} straightforward targeting rules
would reduce its overall PER by about \stat{benchmark.projected.per.reduction}.
The case study also reports that experienced caseworkers spent
under \stat{benchmark.income.minutes.upper} minutes per case
checking only for errors in earned and unearned income.
The state also considered checking other parts of a case,
but timing exercises showed that these broader reviews
would require about \stat{benchmark.expanded.minutes} minutes per case.

\section{Selection approaches and simulation design}
\label{sec:methods}

Under preauthorization review,
after caseworkers have processed an application or renewal,
the case receives an additional check for errors
from an experienced staff member
before benefits are sent out.
Unlike the intensive QC reviews described above,
these checks can be brief and focus on selected parts of a case,
such as earned and unearned income.
If the reviewer finds an error,
the case is returned to the original caseworker for correction.
Correcting errors at this stage prevents incorrect payments
from contributing to the state's official PER~\citep{snap_review_practices}.

But states generally lack the staff capacity
to conduct a pre-authorization audit on every case.
Agencies interested in preauthorization review
must therefore choose which cases to prioritize
and which parts of those cases to examine.
In particular, we simulate and compare two approaches:
``risk-based review,'' where cases are selected
based on the estimated risk of a payment error;
and ``AI-based review,'' where cases are selected
based on specific potential errors flagged by an AI tool.
We also evaluate these approaches in combination.
Next, we describe each method in more detail,
along with the assumptions underlying our simulations.

\subsection{Risk-based review}
\label{sec:risk-review}

\paragraph{Approach.}
Risk-based review uses machine learning models
to predict each case's probability of a payment error,
then orders cases for review from highest to lowest predicted probability.
These models estimate relationships between household characteristics and payment errors
using historical cases that have already been checked for errors.
They then use those relationships to predict error probabilities
for current cases where it is not yet known whether an error is present.
These models help an agency rank cases for review,
but do not identify specific errors;
reviewers must examine the selected cases (in whole or in part)
to determine whether anything needs correction.
Even cases ranked as high-risk may not have any errors at all.

\paragraph{Simulation assumptions.}
In our simulation, we predict risk using household size,
income, deductions from income,
shelter expenses,
the issued benefit amount relative to the maximum benefit,
and time since the household was last approved for benefits.
Because the QC records reflect information gathered during QC review,
we use the source code's reconstruction procedures
to approximate the household information used in the original benefit decision.
We train the model on eligible records from FY\stat{training.years}
and generate predictions for FY\stat{evaluation.year} cases
with complete model inputs.
The outcome we predict is whether the issued benefit
differs from the correct benefit amount
by more than that year's payment error tolerance
(\$\stat{per.tolerance} in FY\stat{evaluation.year}).

Following the BGL SNAP risk model specification,
we use a machine-learning method called ``gradient-boosted decision trees,''
implemented in the XGBoost software package.
We fit each model once using a fixed random seed
and retain the resulting predictions throughout the main simulations.
We fit separate models for households with
\stat{risk.household.size.groups} members;
using separate models
allows the relationships between household characteristics and payment errors
to differ by household size.
Each model pools training cases across the included states and DC
and applies the same prediction function across these jurisdictions.
Then,
for each case in our simulation sample,
the corresponding household-size model produces a predicted error probability.

We use these predicted probabilities
to order cases for simulated review within each state,
starting with the highest-risk case.
These probabilities indicate how likely a case is to contain an error,
but do not tell reviewers \textit{where} to look for errors.
As in the real-world case study described above,
we assume that a state decides in advance
which categories reviewers will check---e.g.,
in the case study, reviewers only checked
earned and unearned income
during preauthorization review.
We call the set of categories checked in each case the ``review scope.''
States may not know in advance which review scope will work best.
To define the scopes we test,
we rank categories by total error dollars in each state's historical QC records.
A scope covering a given number of categories includes
that many of the state's highest-ranked categories.

We test each of these review scopes
with different amounts of staff time.
For example, suppose one review scope includes
the earned and unearned income categories.
Under this review scope,
we start by checking both categories in the highest-risk case.
With more staff time,
we check both categories in the next-highest-risk case,
and continue down the case risk order
until every case in the simulation sample has received these checks.
We then repeat this simulation with different review scopes---for example, adding a check of shelter and utilities to every reviewed case.
Each review scope therefore produces its own sequence of reviews,
from no cases reviewed to all cases reviewed.
We test review scopes ranging from \stat{review.scope.min.words} category
through all \stat{review.scope.max.words} categories.

We assume that each category check takes \stat{review.risk.minutes} minutes
and has a \stat{review.risk.correction} probability
of finding and correcting all errors in the part of the case being checked.
(Our conclusions are largely stable
with different review times and higher correction success;
see Appendices~\ref{apx:sensitivity-category-time}
and~\ref{apx:sensitivity-category-correction}.)
For example, a successful earned-income check
corrects all errors in earned income;
an unsuccessful check leaves those errors unchanged.
We assume that success on one check is independent of success on another.
The time required is the same whether or not the check finds an error.
We hold our time and correction assumptions fixed across review scopes,
so expanding review scope linearly increases the time spent on each case.

\subsection{AI-based review}
\label{sec:ai-review}

\paragraph{Approach.}
We next consider the use of an AI flagging system
that would identify potential payment errors for preauthorization review.
The design we consider is inspired by existing uses of large language models
to process unstructured government records,
including redacting race-related information from police reports
and identifying racial covenants in property records
for legal review~\citep{walia2026validation,surani2026covenants}.
In the system we envision,
after a caseworker makes a preliminary eligibility and benefit determination,
a large language model would review the documents in the case file.
It would check whether the documents,
information entered by the caseworker, and calculations agree,
and identify implausible or inconsistent information.
It would also check whether the eligibility determination and calculated benefit amount
are correct given the information available in the case,
as defined by state policy in written instructions given to caseworkers.

If the system detects any such discrepancies,
it would flag them for a reviewer to check.
Each flag would show a reviewer the specific potential discrepancy
and the information needed to check it.
An experienced reviewer could then assess the flagged issue directly.
If the reviewer confirms an error,
the case would be returned to the original caseworker for correction;
otherwise, the flag would be dismissed.

\paragraph{Simulation assumptions.}
Notably, the system (at least as envisioned above)
would only be able to flag errors visible
in the information already reported by the applicant.
To represent this limitation in our simulation,
we use the ``primary cause of variance''
to define which errors we assume AI can detect.
For example, we assume that AI can detect data-entry errors
(e.g., incorrectly entered wages from a submitted pay stub)
because it could compare the entered amount with the source document.
Similarly, we assume it cannot detect unreported income
(e.g., income from a job the applicant never disclosed)
because case documents would be unlikely to have any indication 
that the income is missing.
Appendix~\ref{apx:ai-error-coverage} lists all causes of variance
and whether we assume AI can detect the associated errors.

We construct simulated AI flags by perturbing the detectable errors
recorded in the public-use file,
randomly omitting some and randomly adding flags for errors that do not exist.
Specifically, we assume each detectable error
has a \stat{review.ai.recall} probability of generating a flag,
independently of other errors
(see Appendix~\ref{apx:sensitivity-detection} for sensitivity to this probability).
Within each state and category,
we set the expected number of false flags
so that \stat{review.ai.precision} of flags correspond to actual errors.

The flags we simulate identify potential problems 
but do not establish a natural review order.
When there is not enough staff capacity to check every flag,
the agency must decide which to review first.
Similar to the risk-based approach,
we assume reviewers concentrate their time on flags in the top-ranked categories
(e.g., earned income or shelter and utilities),
ranked by total error dollars in that state's FY\stat{training.years} records.
In our simulation, reviewers start with the category with the largest historical total
and check all its flags before moving to the next-highest category,
reviewing flags within each category in random order.
(Our conclusions are largely unchanged when we use other review orders,
such as prioritizing categories by estimated error reduction per flag
or reviewing all flags in random order;
see Appendix~\ref{apx:sensitivity-ai-order}.)

We assume that reviewing each flag takes \stat{review.ai.minutes} minutes,
whether or not it identifies a real error.
Even when AI correctly flags a real error,
human review may fail to confirm or correct it.
We assume that human review results in correction
with \stat{review.ai.correction} probability for each real error reviewed,
independently of other errors.
We also assume that a successful review corrects the specific error flagged,
rather than all errors in that category of the case.
With more staff time,
reviewers work through more flags in the order described above,
until all flags have been checked.
We report sensitivity to review time and correction success
in Appendices~\ref{apx:sensitivity-flag-time}
and~\ref{apx:sensitivity-flag-correction}, respectively.

\paragraph{Comparing the approaches.}
\label{sec:comparison}
The AI-based approach has both advantages and drawbacks
compared to the risk-based approach,
making it unclear which would perform better in practice.
Risk-based review directs staff toward cases most likely to contain errors,
but reviewers must search for problems that may or may not exist.
At the same time, high risk for a given case
can indicate that there are potential errors of omission
that are not apparent in the submitted documents.
On the other hand,
AI-based review screens \textit{all} cases,
regardless of their risk ranking,
and can point staff directly to a specific problem,
which promises to reduce the time needed to check each problem.
But it cannot flag missing information
that leaves no trace in the case documents---such as income from an undisclosed job---so errors caused by these omissions would remain unflagged
and thus uncorrected.

\subsection{Combined review}
\label{sec:combined-review}

Finally, we simulate how risk scores and AI flags could be used together
in preauthorization review.
These approaches are promising in combination
because risk scores provide a way to prioritize cases,
while AI flags could quickly direct reviewers to specific potential problems within them.
In our simulation, we assume reviewers first check all AI flags,
in descending order of case risk as rated by the risk model.
After all AI flags have been checked,
reviewers turn to cases without flags,
following the same risk ordering and category checks
as in the standalone risk-based approach.
In addition, we assume that cases with flags receive only flag checks;
reviewers do not subsequently search those cases for unflagged errors.
This combined approach retains the time and correction assumptions described above.
(We also simulate an alternative combined approach
that follows the risk-based order for all cases,
using flag checks in flagged categories and ordinary reviews otherwise.
This alternative generally leaves more payment errors uncorrected
at a fixed amount of staff time;
see Appendix~\ref{apx:sensitivity-combined-order}.)

\subsection{Other assumptions}
\label{sec:other-assumptions}

For standalone risk-based and AI-based review,
we enumerate all possible combinations of successful and unsuccessful corrections
for each household.
We weight each combination by its probability
to calculate expected error reduction,
and calculate expected review time from the number of category checks or flags.
For combined review, flags also determine whether a household
receives flag checks or ordinary category reviews.
We estimate expected outcomes by simulating \stat{simulation.combined.replicates}
sets of flags and correction outcomes for each household,
using a fixed random seed.
In these draws, each detectable error is independently flagged
and each reviewed error or category is independently corrected
with the probabilities described above.
We draw false-flag counts independently from a Poisson distribution,
with the same mean across households within each state and category;
households with no errors can also receive false flags.
We average review times and error reductions across these draws
before constructing review queues,
and then finally use QC sampling weights to obtain state totals.
In all approaches, we apply the payment-error tolerance
to individual errors in each outcome before averaging.

We convert review time into labor costs using
\$\stat{labor.hourly.total} per hour for all approaches,
a rounded assumption that includes wages and benefits.
For context, the Bureau of Labor Statistics reports a national mean hourly wage
of \$\stat{labor.benchmark.wage}
for eligibility interviewers in government programs
in \stat{labor.benchmark.year}~\citep{bls_eligibility_wages}.
This is a benchmark for the occupation as a whole;
experienced reviewers likely earn more.
We assume that the additional reviews qualify for federal reimbursement
at \stat{labor.federal.share},
the federal share of SNAP administrative costs beginning in
FY\stat{labor.share.year}~\citep{snap_admin_cost_sharing}.
After accounting for this reimbursement,
the resulting state labor cost is \$\stat{labor.hourly.state} per hour.
For simplicity, we exclude technical development and integration costs,
which we treat as one-time costs of establishing each approach,
and focus on the recurring labor needed to review cases.
We also exclude the time spent correcting errors,
assuming that correcting a given error costs the same
regardless of how it was identified.
Our labor-cost comparisons therefore measure review costs,
rather than the full costs of implementing each approach.

\section{Results}
\label{sec:results}

We evaluate the three approaches above
by simulating how much payment error remains
as state administrators devote increasing amounts of staff capacity
to preauthorization review.
Our central question is whether these approaches can bring a state
below the \stat{per.cost.sharing.threshold}\% PER threshold
and, where that is possible,
which approach does so at the lowest review cost to the state.
Reaching this threshold would allow the state to avoid benefit cost sharing,
which can amount to hundreds of millions of dollars per year for an individual state.
As described above,
for risk-based and combined review,
we simulate different review scopes
(e.g., checking only earned and unearned income).
At each review cost,
we report the largest reduction in PER across the review scopes we test,
selecting the best-performing scope with the benefit of hindsight.
This gives an optimistic estimate of what agencies could achieve in practice,
since an agency would have to choose a single review scope
for preauthorization review
without knowing in advance which would perform best.

\begin{figure}[t]
\centering
\includegraphics[width=\linewidth]{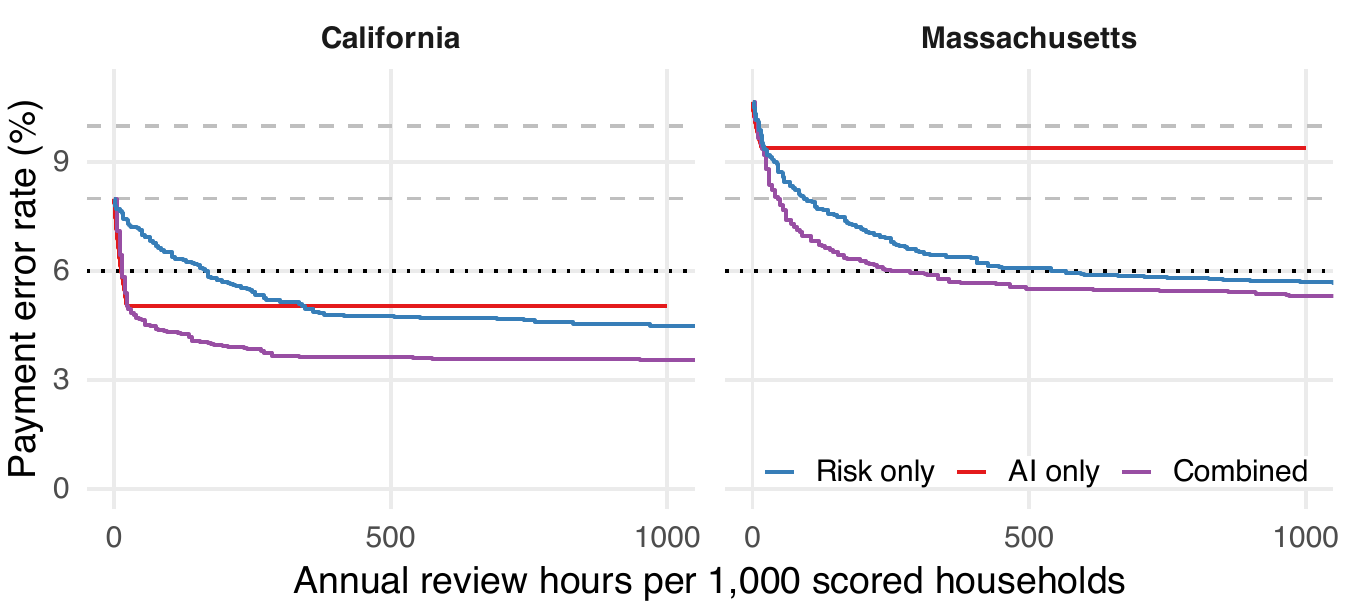}
\caption{
Review capacity vs. payment error rates in California and Massachusetts.
The horizontal axis shows annual reviewer time relative to the scored caseload;
the vertical axis shows the remaining statewide PER.
A lower curve indicates fewer payment errors for the same amount of staff time.
The black dotted line marks the \stat{per.cost.sharing.threshold}\% PER threshold
below which states avoid benefit cost sharing.
AI-based review reaches this threshold with little staff time in California
but cannot reach it in Massachusetts.
Combined review reaches the threshold in both states
with substantially less staff time than risk-based review alone.
}
\label{fig:review-capacity}
\end{figure}

We first focus on California and Massachusetts,
which illustrate different patterns of error reduction
(Figure~\ref{fig:review-capacity}).
In California, AI-based review quickly brings PER
below the \stat{per.cost.sharing.threshold}\% threshold,
while risk-based review requires about
\stat{results.ca.risk.ai.time.ratio} times as much staff time to reach that target.
In Massachusetts, on the other hand,
AI-based review produces only a modest reduction in PER
before exhausting the errors it can detect,
leaving the state above the cost-sharing threshold
and thus liable to hundreds of millions in cost-sharing penalties.
Risk-based review can reduce PER further and reach the target,
but requires more extensive review.
In contrast,
the combined approach produces lower PER than either approach alone
over much of the range of staff capacity we examine.
In both states,
combined review reaches the threshold with substantially less staff time
than risk-based review alone,
which costs approximately
\stat{results.risk.combined.cost.ratio.range} times as much
as combined review to reach the cost-sharing penalty threshold.

\begin{figure}[t]
\centering
\includegraphics[width=\linewidth]{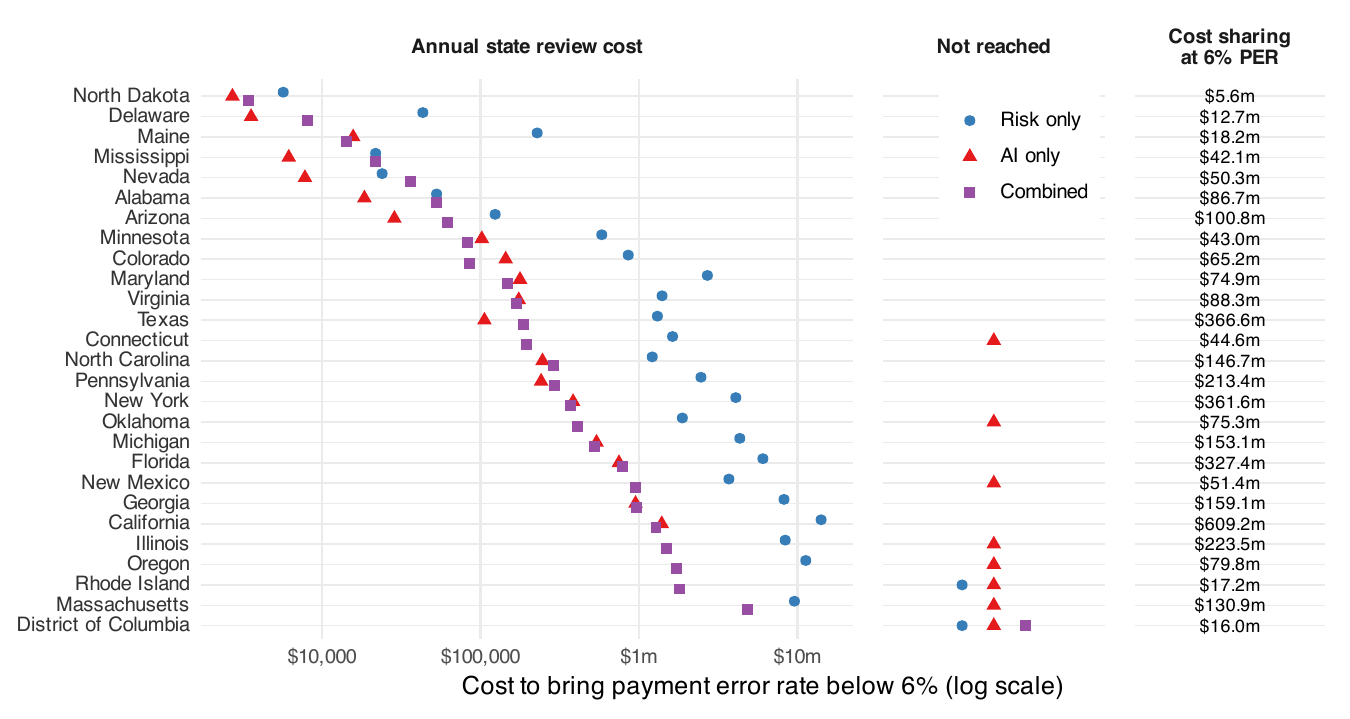}
\caption{
Minimum annual state review cost to bring PER
below \stat{per.cost.sharing.threshold}\%.
The figure includes the \stat{results.above.threshold.count} jurisdictions
that begin at or above this threshold,
ordered by the cost of combined review.
Symbols show the lowest review cost among the simulated designs
that reach the target, on a logarithmic scale.
``Not reached'' indicates that no tested design for that approach
brings the jurisdiction below the threshold.
Costs reflect the state's share of review labor costs.
The rightmost column shows annual benefit cost sharing
at the lowest nonzero rate, \stat{benefit.share.low}\% of annual benefits,
which applies at a PER just above \stat{per.cost.sharing.threshold}\%.
Combined review reaches the target in more jurisdictions than either approach alone
and generally costs less than risk-based review.
}
\label{fig:state-review-costs}
\end{figure}

To compare results across jurisdictions,
we focus on the \stat{results.above.threshold.count}
that begin at or above the \stat{per.cost.sharing.threshold}\% PER threshold
(Figure~\ref{fig:state-review-costs}).
(Other jurisdictions
are already below the PER threshold
and would not need additional review to reach it.)
Combined review brings \stat{results.combined.target.count}
of these \stat{results.above.threshold.count} jurisdictions below the threshold,
compared with \stat{results.risk.target.count} for risk-based review
and \stat{results.ai.target.count} for AI-based review.
None of the three main approaches we simulate bring the District of Columbia
below the threshold within the review budgets we test.
Among the \stat{results.shared.target.count} jurisdictions
where both risk-based and combined review reach the target,
combined review requires less spending on reviews
in \stat{results.combined.cheaper.count}.
At the median across all \stat{results.shared.target.count} jurisdictions
where both approaches reach the target,
the risk-based approach costs about
\stat{results.risk.combined.cost.ratio.median} times as much
as the combined approach.

The large cost differences between the combined approach
and the risk-based review approach
reflect both how reviews are targeted
and our assumptions about the work required.
In our simulation, checking an AI flag takes less time
and has a higher chance of correcting a real error
than an ordinary category review,
which requires reviewers to search for errors that may not exist.
Combined review can further reduce costs by checking higher-risk flagged cases first
and then later searching unflagged cases for errors that AI misses.
But prioritizing cases by risk is not always the cheapest way to reach the target:
in some states (i.e., Delaware, Mississippi, Nevada, Alabama, and Arizona),
AI-only review can cost less by prioritizing review of historically costly error categories.
D.C. is a near miss: reaching the threshold requires eliminating about
\stat{results.dc.required.reduction}\% of its initial error dollars,
but combined review eliminates only
\stat{results.dc.achieved.reduction}\% of error dollars.
In D.C., our simulated preauthorization reviews cannot close this gap entirely,
because flagged cases receive no further search for errors,
some detections fail,
and cases without risk scores remain outside the review population.
Notably, across all scenarios we model,
the review costs required to reach the target
are generally \textit{much} smaller than even the lowest cost-sharing penalty
that a state would face at or above \stat{per.cost.sharing.threshold}\% PER
(Figure~\ref{fig:state-review-costs}).

These findings are largely robust
to a wide range of plausible alternate simulation assumptions
(Figure~\ref{fig:sensitivity-overview}; Appendix~\ref{apx:sensitivity}).
The largest changes we observe concern how many jurisdictions
AI-only review can bring below the cost-sharing threshold:
varying the types of errors that AI is able to detect changes this number
from \stat{sensitivity.scope.narrow.targets}
to \stat{sensitivity.scope.broad.targets}
(Appendix~\ref{apx:sensitivity-coverage}).
Nevertheless, in every sensitivity scenario---including faster or slower reviews,
much stronger risk-model performance (Appendix~\ref{apx:sensitivity-risk-scores}),
higher or lower AI detection rates,
changes in reviewers' correction success under both risk-based and AI-based review,
and narrower or broader AI error coverage---combined review costs less than risk-only review
in most jurisdictions where both reach the target.

\section{Discussion}
\label{sec:discussion}

Our simulations suggest that combining risk scores with AI-generated flags
could help states reduce SNAP payment errors
at substantially lower review costs than risk-based review alone.
Under our main assumptions, combined review allows
\stat{results.combined.target.count} of the \stat{results.above.threshold.count}
jurisdictions facing cost-sharing penalties to avoid them,
at preauthorization review costs
that are a fraction of those threatened penalties.
The combined approach uses risk scores to prioritize cases,
AI flags to direct reviewers to potential discrepancies,
and ordinary review to find errors that AI misses in unflagged cases.
Its cost advantage over risk-based review persists across a wide range of assumptions---including substantially more accurate risk scores---suggesting that the two tools could serve complementary roles in preauthorization review.

Payment accuracy is only one measure of how well SNAP serves eligible households.
Ideally, these tools would also help people receive the correct benefits sooner,
by reducing the time that staff take to check cases
and by quickly resolving discrepancies before they delay the provision of benefits.
But preauthorization review also runs the risk of slowing approval
or could lead to burdensome requests for additional documents,
especially when AI flags are incorrect or difficult to resolve.
Our simulations do not measure these effects,
nor do they establish whether such tools would reduce erroneous denials
or improve participation among eligible households.
Evaluations in practice should therefore measure processing times,
applicants' reporting and documentation burdens,
and access to the correct benefit amount,
alongside payment error rates and review costs.

Whether AI can identify SNAP errors reliably enough
to deliver these gains remains an open question.
We simulate under a set of assumptions that seem plausible to us
about detection and correction,
but performance may be different in practice.
Testing this approach would require access to a large number of case documents,
a state's written policy instructions,
and independent assessments of which cases contain errors.
Initial tests could compare AI flags with human assessments,
followed by pilots measuring how accurately and quickly
reviewers resolve flags in their normal workflow.
States will likely need technical assistance to 
develop such tools,
assemble evaluation data, 
integrate tools into existing systems,
and train staff to use them.
BGL's open-source risk model~\citep{snap_risk_code}, which we use in this study,
provides one example of how open-source software can accelerate
the development and evaluation of tools for public agencies.
The Safety Net Response Network also helps states share practical approaches
to reducing payment errors~\citep{snap_review_practices}.
Similar open efforts could accelerate the development of AI-based review tools.

By threatening states with potential cost-sharing penalties 
if they fail to reduce PER below 6\%,
H.R.~1 could make access to food assistance 
more dependent on states' willingness to fund it.
Temporary Assistance for Needy Families (TANF),
which has long required states to pay for a portion of costs,
provides a warning:
over time, states have directed more resources away from TANF,
and fewer eligible families have
received direct cash assistance as a result~\citep{gao_tanf_2013}.
On the other hand,
if the gains we simulate can be achieved in practice,
AI could help states reduce costs while potentially expanding access to benefits.
Achieving this outcome will require states to deploy AI
that helps reviewers identify and resolve payment errors more quickly.
Investing in that capacity could help states protect public benefits
and deliver them more accurately and promptly to the people who need them.

\section*{Acknowledgements}
I am grateful to Ravi Shroff for his thoughtful and insightful comments
on an early draft of this analysis.

\begingroup
\raggedright
\bibliographystyle{unsrtnat}
\bibliography{references}
\endgroup

\clearpage
\appendix
\setcounter{figure}{0}
\renewcommand{\thefigure}{\thesection\arabic{figure}}
\renewcommand{\theHfigure}{appendix.\thesection.\arabic{figure}}
\renewcommand{\thetable}{\thesection\arabic{table}}
\section{Public-use and official payment error rates}
\label{apx:per-comparison}

\begin{figure}[!htbp]
\centering
\includegraphics[width=0.85\linewidth]{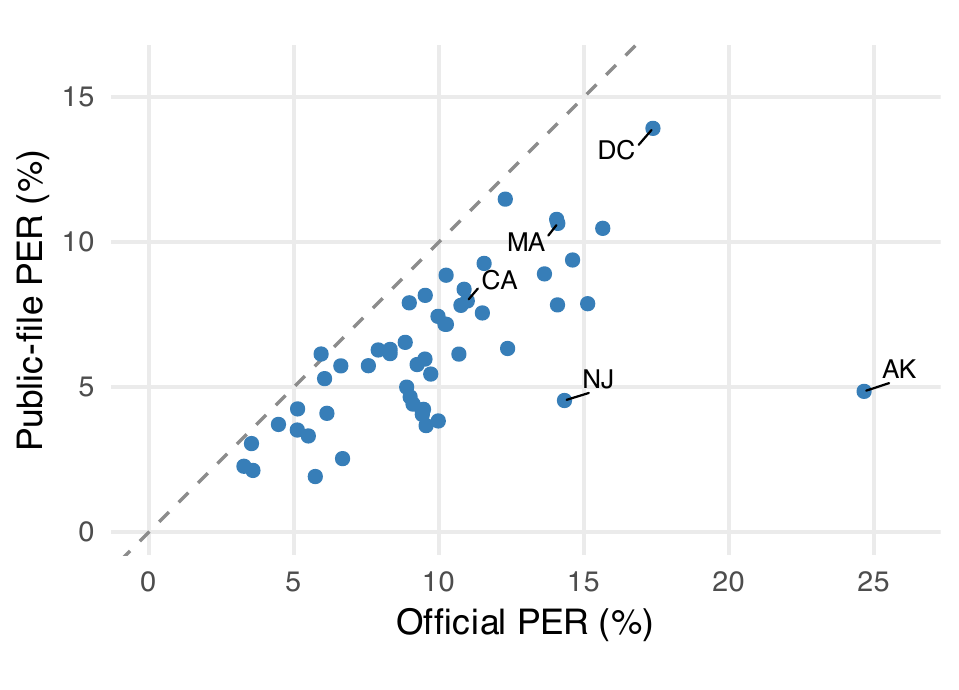}
\caption{Public-file PER tracks official PER, but is generally lower.
Official and public-file payment error rates by jurisdiction,
FY\stat{evaluation.year}.
Each point represents a state, the District of Columbia, Guam,
or the Virgin Islands; the dashed line indicates equality.
Labels identify our example states, DC, and the largest discrepancies.
Sources: USDA's official PER table and SNAP QC public-use file~\citep{snap_official_per,snap_qc_data}.}
\label{fig:per-comparison}
\end{figure}

Figure~\ref{fig:per-comparison} compares the payment error rates
we calculate from the FY\stat{evaluation.year} public-use QC file
with USDA's official rates for the same year.
For each jurisdiction, we divide survey-weighted error dollars
by survey-weighted benefit dollars,
counting only case errors above the \$\stat{per.tolerance} reporting tolerance.
The rates are strongly correlated among jurisdictions included in our simulations
(Pearson correlation \stat{apx.per.sim.correlation});
across the full public file, the correlation is \stat{apx.per.correlation}.
Public-file rates are lower in \stat{apx.per.lower}
of the \stat{apx.per.count} jurisdictions shown.
The public file thus broadly tracks official PER rates,
but does not reproduce the official rates exactly.
Our simulations use the public-file rates throughout,
so the review costs we estimate apply to those starting rates.

\clearpage
\setcounter{table}{0}
\section{AI error coverage}
\label{apx:ai-error-coverage}

Table~\ref{tab:ai-coverage} lists the QC file's primary causes of variance
and our assumptions about whether AI could flag the associated errors.
The descriptions and codes come from the QC documentation;
the detectability assignments are ours.
We assume that discrepancies in reported information,
calculations, or required procedures can be identified from the case record,
but that unreported information generally cannot.
These are broad assumptions about error types,
not judgments about the visibility of each individual error.
In particular, we treat the mixed category of incomplete or incorrect information
as detectable, even though some errors in that category may leave no visible discrepancy.
Detectable errors still generate flags only with the probability specified
in Section~\ref{sec:ai-review}.

\begin{table}[htbp]
\centering
\caption{Assumed AI detectability by primary cause of variance.
Codes and descriptions follow the SNAP QC documentation~\citep{snap_qc_data}.
Missing causes are treated as detectable in the main simulation.}
\label{tab:ai-coverage}
\begingroup
\footnotesize
\setstretch{1.0}
\setlength{\tabcolsep}{5pt}
\input{ai_coverage_table.tex}
\endgroup
\end{table}

\clearpage
\setcounter{table}{0}
\section{Review categories}
\label{apx:review-categories}

Table~\ref{tab:review-categories} shows how we group the QC file's
elements of variance into categories for simulated review.
Each category represents related parts of a case that a reviewer could check together.
The groupings are ours;
the element codes and descriptions come from the QC documentation~\citep{snap_qc_data}.

\begingroup
\small
\setstretch{1.0}
\setlength{\tabcolsep}{5pt}
\begin{longtable}{@{}>{\raggedright\arraybackslash}p{0.23\linewidth}>{\raggedright\arraybackslash}p{0.72\linewidth}@{}}
\caption{Review categories and their constituent elements of variance.}
\label{tab:review-categories}\\
\toprule
\textbf{Review category} & \textbf{QC elements of variance} \\
\midrule
\endfirsthead
\multicolumn{2}{l}{\textit{Table~\ref{tab:review-categories}, continued}} \\
\toprule
\textbf{Review category} & \textbf{QC elements of variance} \\
\midrule
\endhead
\midrule
\multicolumn{2}{r}{\textit{Continued on the next page}} \\
\endfoot
\bottomrule
\endlastfoot
\input{review_categories_table.tex}
\end{longtable}
\endgroup

\clearpage
\setcounter{figure}{0}
\section{Sensitivity analysis}
\label{apx:sensitivity}

\begin{figure}[!htbp]
\centering
\includegraphics[width=\linewidth]{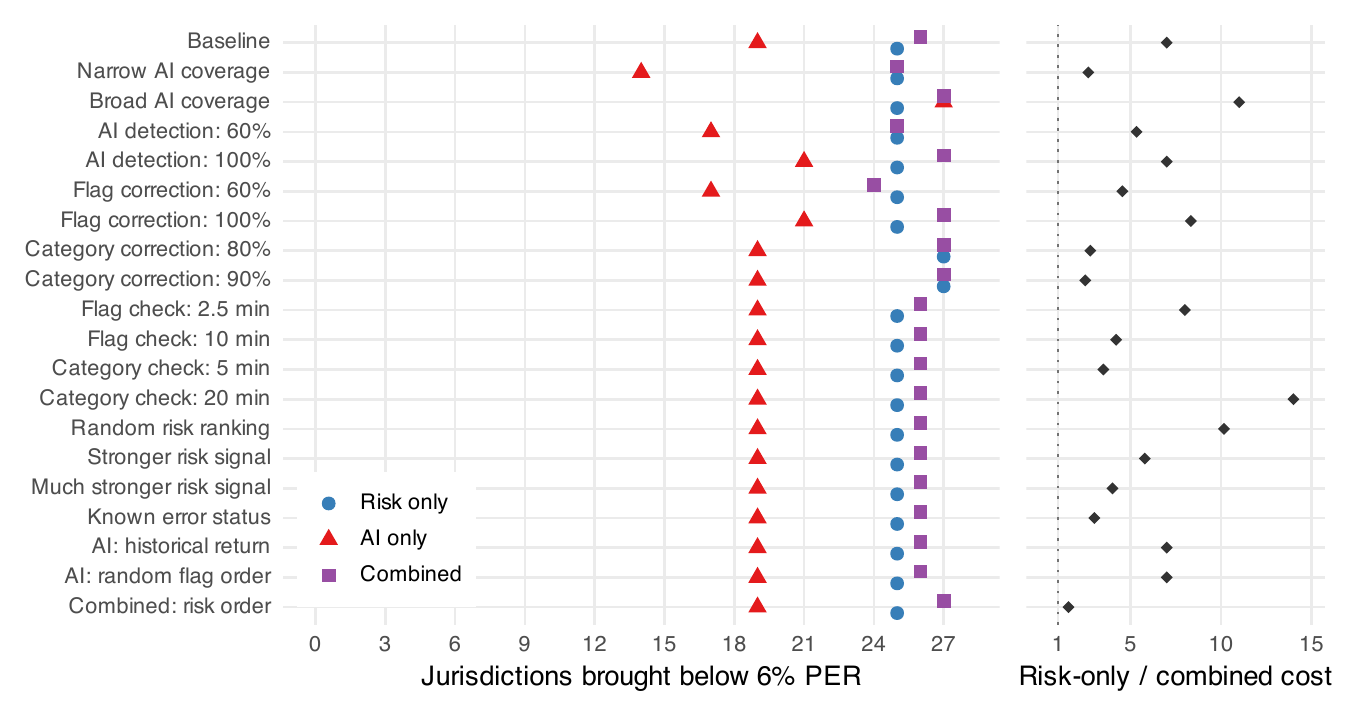}
\caption{Sensitivity overview. Left: jurisdictions brought below the threshold.
Right: median within-jurisdiction ratio of risk-only to combined review costs,
among jurisdictions where both reach the target; values above one favor combined review.
For repeated risk-score draws, we first take each jurisdiction's median cost.}
\label{fig:sensitivity-overview}
\end{figure}

We vary the assumptions below one at a time,
holding the remaining assumptions at their main values.
Figure~\ref{fig:sensitivity-overview} shows how many of the
\stat{results.above.threshold.count} initially above-threshold jurisdictions
each approach brings below \stat{per.cost.sharing.threshold}\% PER.

\begingroup
\setlength{\parskip}{0pt}
\begin{enumerate}
\setstretch{1.0}
\setlength{\parsep}{0pt}
\setlength{\itemsep}{0pt}
\item AI error coverage: narrower and broader sets of detectable errors.
\item AI detection: lower and higher probabilities of flagging an error.
\item Human flag correction: lower and higher success rates.
\item Ordinary category review: higher human correction success rates.
\item Flag-review time: shorter and longer checks.
\item Category-review time: shorter and longer checks.
\item Risk scores: random rankings, stronger signals, and known error status.
\item AI flag order: historical correction per hour and random ordering.
\item Combined review: following case risk throughout rather than flags first.
\end{enumerate}
\endgroup

\clearpage
\subsection{Which errors AI can detect}
\label{apx:sensitivity-coverage}

The narrow-coverage scenario includes policy, data-entry, and calculation errors
(codes 10, 12, and 17--22 in Table~\ref{tab:ai-coverage}),
but excludes incomplete information, follow-up, verification,
recertification, and missing causes.
Broad coverage adds unreported, withheld, or inaccurate information
from applicants and collateral contacts (codes 1--4 and 7)
to the main detectable set.
It assumes access to the information needed to identify these errors,
even when absent from the submitted documents.
Codes 8, 26, and 99 remain excluded.

AI-only review reaches \stat{apx.targets.ai_scope_narrow.ai} jurisdictions
under narrow coverage and \stat{apx.targets.ai_scope_broad.ai} under broad coverage,
compared with \stat{results.ai.target.count} under the main assumptions.
Combined review reaches \stat{apx.targets.ai_scope_narrow.combined}
and \stat{apx.targets.ai_scope_broad.combined}, respectively;
risk-only results are unchanged.
Coverage thus changes what AI can achieve on its own,
while combined review retains ordinary review in unflagged cases.

Here and below, costs exclude remaining benefit cost sharing,
and ``not reached'' means no simulated budget achieves the target.
As in Figure~\ref{fig:state-review-costs},
we select the best-performing review scope at each budget
for risk-only and combined review.

\begin{figure}[!htbp]
\centering
\includegraphics[width=\linewidth]{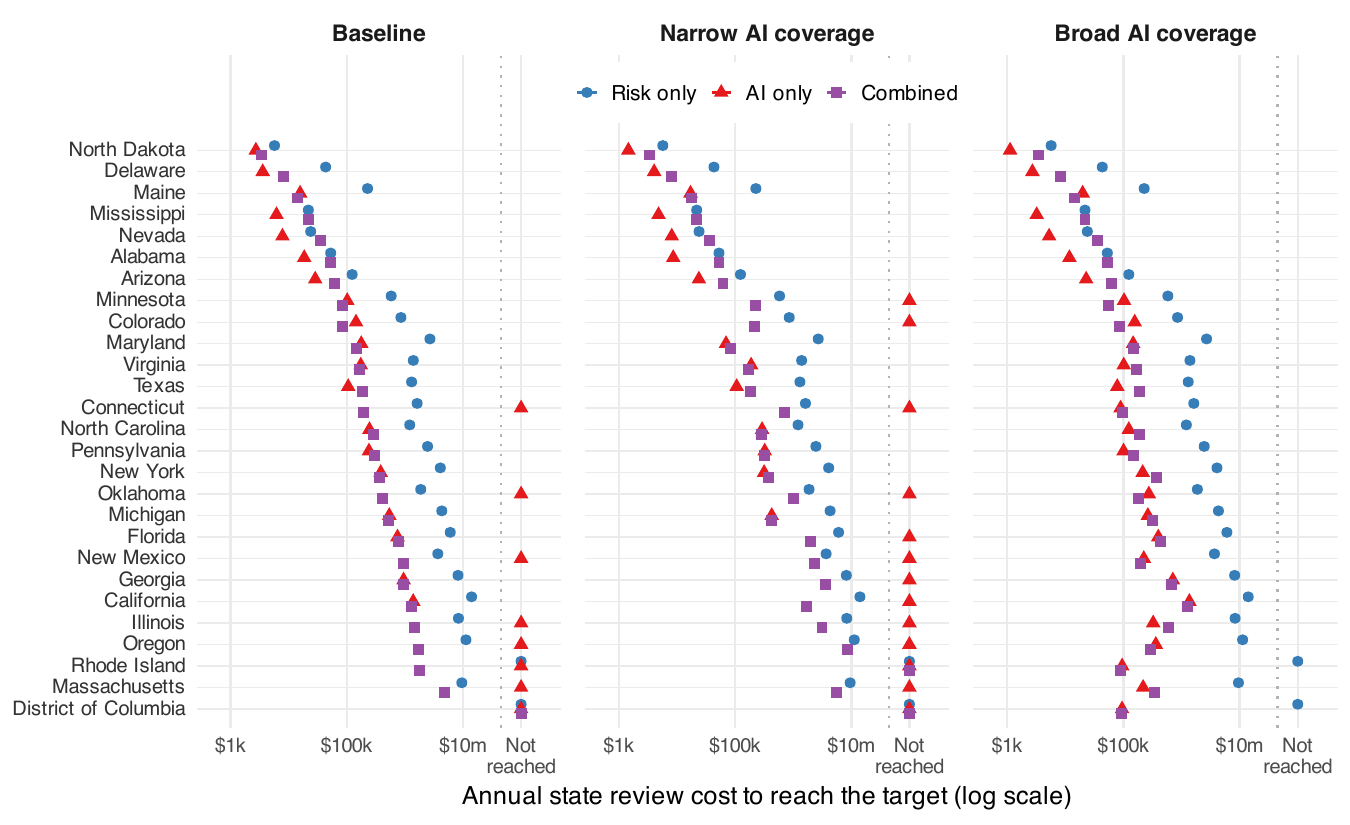}
\caption{Review costs under alternative AI error coverage assumptions.
All panels use the same jurisdictions, ordering, and cost scale.
Points to the right of the dotted divider indicate that the target is not reached.}
\label{fig:sensitivity-coverage}
\end{figure}

\clearpage
\subsection{How reliably AI flags detectable errors}
\label{apx:sensitivity-detection}

We change the probability that a detectable error generates a flag
from \stat{review.ai.recall} to \stat{sensitivity.success.low}\%
or \stat{sensitivity.success.high}\%,
while keeping human correction success at \stat{review.ai.correction}.
The share of flags corresponding to real errors remains
\stat{review.ai.precision}; consequently,
more detected errors also produce more false flags to review.
We do not separately vary this share.

Lower detection reduces the number of jurisdictions reaching the target
to \stat{apx.targets.ai_recall_low.ai} for AI-only review
and \stat{apx.targets.ai_recall_low.combined} for combined review.
With perfect detection of the assumed detectable errors,
these counts rise to \stat{apx.targets.ai_recall_high.ai}
and \stat{apx.targets.ai_recall_high.combined}.
Perfect detection does not make otherwise invisible error types detectable.
The combined approach can use ordinary review to find errors in unflagged cases,
so its reach is less dependent on AI detection alone.

\begin{figure}[!htbp]
\centering
\includegraphics[width=\linewidth]{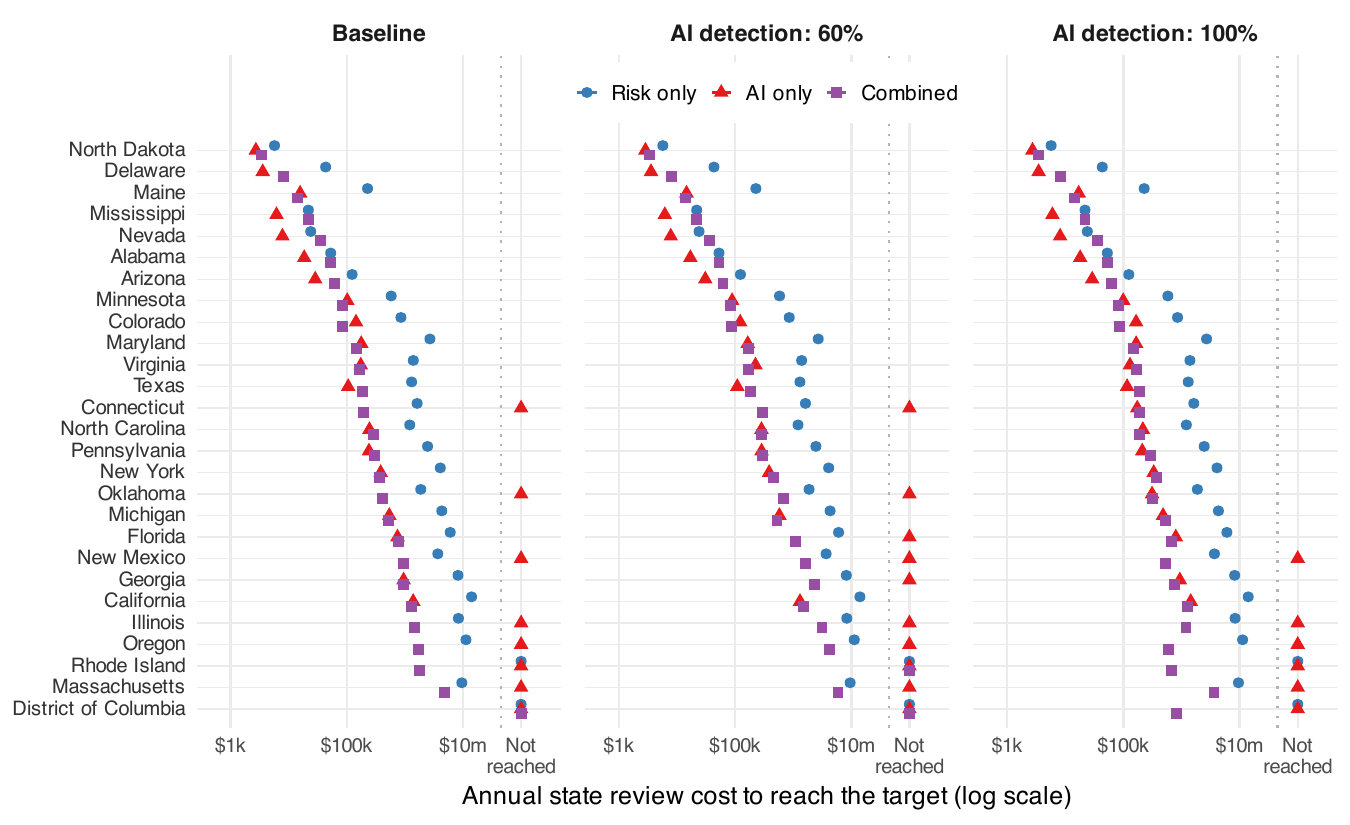}
\caption{Review costs when AI flags a smaller or larger share of detectable errors.
Human correction success and the share of flags that are real remain fixed.}
\label{fig:sensitivity-detection}
\end{figure}

\clearpage
\subsection{Human correction of flagged errors}
\label{apx:sensitivity-flag-correction}

We vary the probability that a reviewer successfully corrects a real flagged error
from \stat{review.ai.correction} to \stat{sensitivity.success.low}\%
or \stat{sensitivity.success.high}\%,
while holding AI detection at \stat{review.ai.recall}.
This check concerns human review after a flag has been generated,
not AI's ability to identify the error.
Failed corrections still consume review time.

At the lower success rate, AI-only review reaches
\stat{apx.targets.ai_success_low.ai} jurisdictions and combined review reaches
\stat{apx.targets.ai_success_low.combined}.
At the higher rate, they reach \stat{apx.targets.ai_success_high.ai}
and \stat{apx.targets.ai_success_high.combined}.
Combined review is more vulnerable to unsuccessful flag correction
than to missed flags: cases with flags receive no subsequent ordinary review,
whereas cases without flags remain eligible for it.

\begin{figure}[!htbp]
\centering
\includegraphics[width=\linewidth]{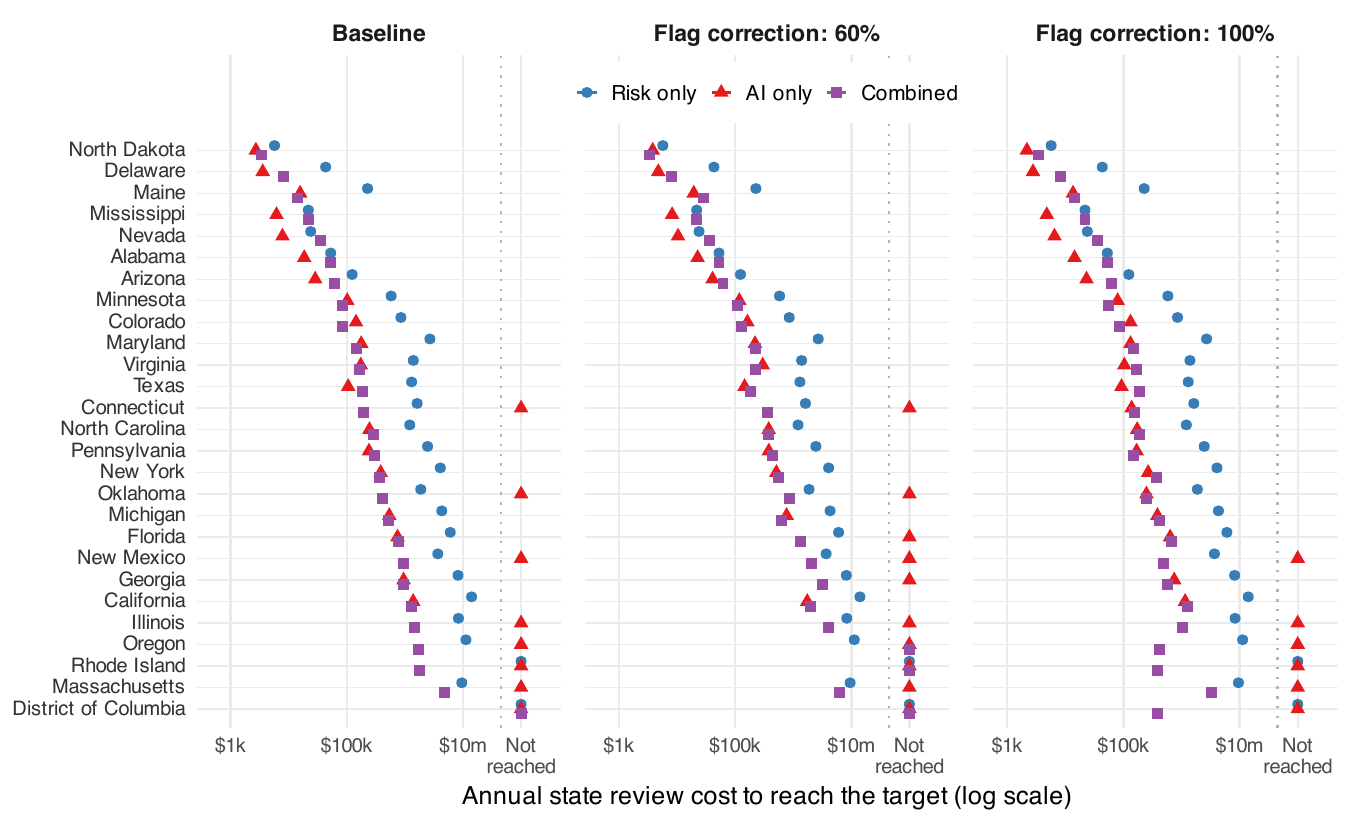}
\caption{Review costs under lower and higher human success in correcting flagged errors.
The probability that AI detects an error remains unchanged.}
\label{fig:sensitivity-flag-correction}
\end{figure}

\clearpage
\subsection{Human correction during ordinary category review}
\label{apx:sensitivity-category-correction}

We increase the probability that an ordinary category check succeeds
from \stat{review.risk.correction} to \stat{apx.category.success.low}\%
or \stat{apx.category.success.high}\%.
This change applies to risk-only review and to the ordinary-review phase
of the combined approach; AI-only review is unchanged.
A successful check corrects all errors in the reviewed category,
including errors outside AI's detectable set.

At either higher success rate, both risk-only and combined review can bring
all \stat{results.above.threshold.count} jurisdictions below the threshold.
More effective ordinary review narrows the cost difference:
risk-only review costs a median of
\stat{apx.ratio.risk_success_high} times as much as combined review
at \stat{apx.category.success.low}\% success,
and \stat{apx.ratio.risk_success_90} times as much
at \stat{apx.category.success.high}\% success.
These medians compare costs within jurisdictions where both methods reach the target.

\begin{figure}[!htbp]
\centering
\includegraphics[width=\linewidth]{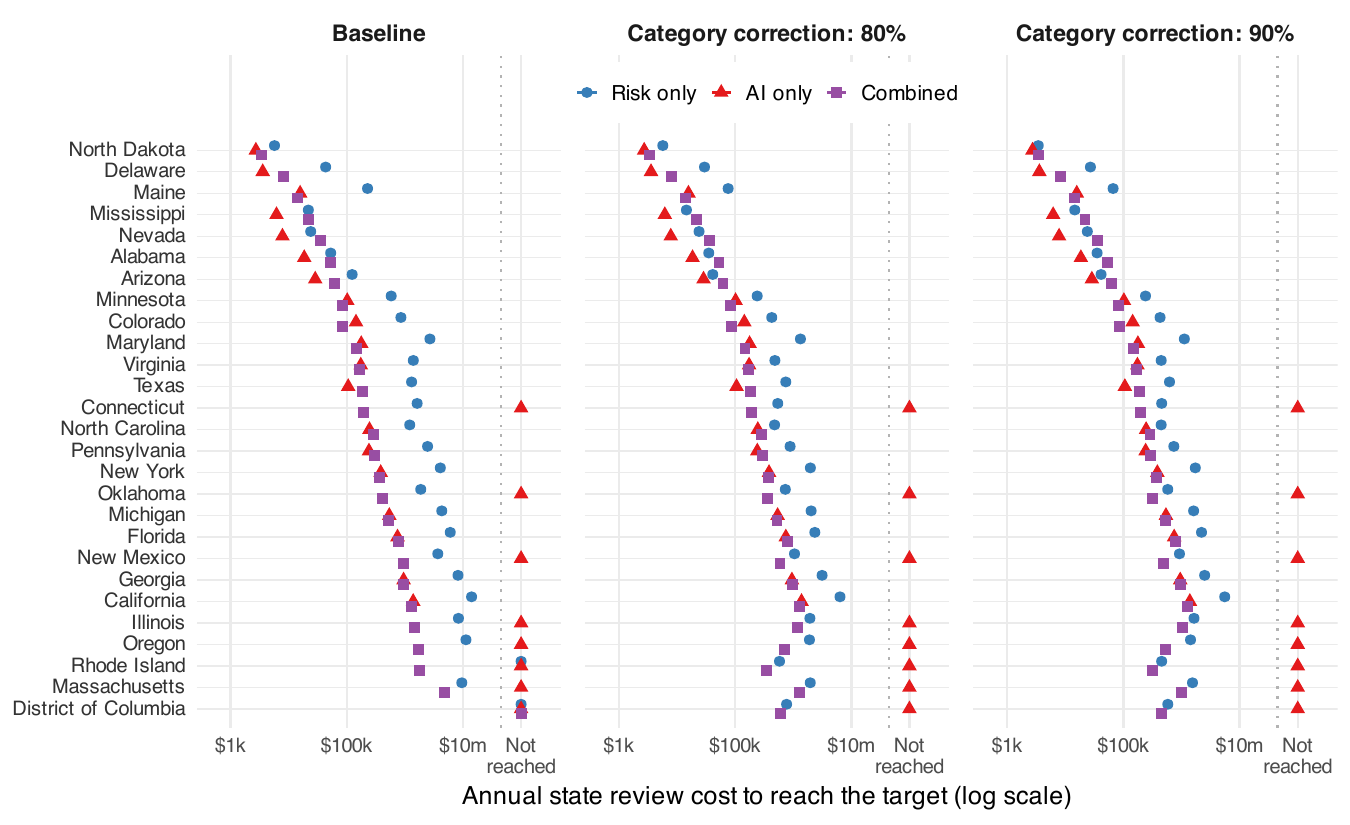}
\caption{Review costs when ordinary category checks are more effective.
Both approaches using ordinary review reach more jurisdictions,
but combined review generally remains less costly.}
\label{fig:sensitivity-category-correction}
\end{figure}

\clearpage
\subsection{Time spent checking an AI flag}
\label{apx:sensitivity-flag-time}

We change review time per flag from \stat{review.ai.minutes} minutes
to \stat{apx.flag.time.low} or \stat{apx.flag.time.high} minutes.
The time applies equally to true and false flags.
Changing review time affects costs without changing which errors can be corrected,
so each approach reaches the same jurisdictions as in the main simulation.
AI-only costs scale directly with time per flag.
Combined costs also change, but only for the part of the work spent checking flags.
Among jurisdictions where both reach the target,
risk-only review costs a median of \stat{apx.ratio.ai_time_low} times as much
as combined review with shorter flag checks,
and \stat{apx.ratio.ai_time_high} times as much with longer checks.

\begin{figure}[!htbp]
\centering
\includegraphics[width=\linewidth]{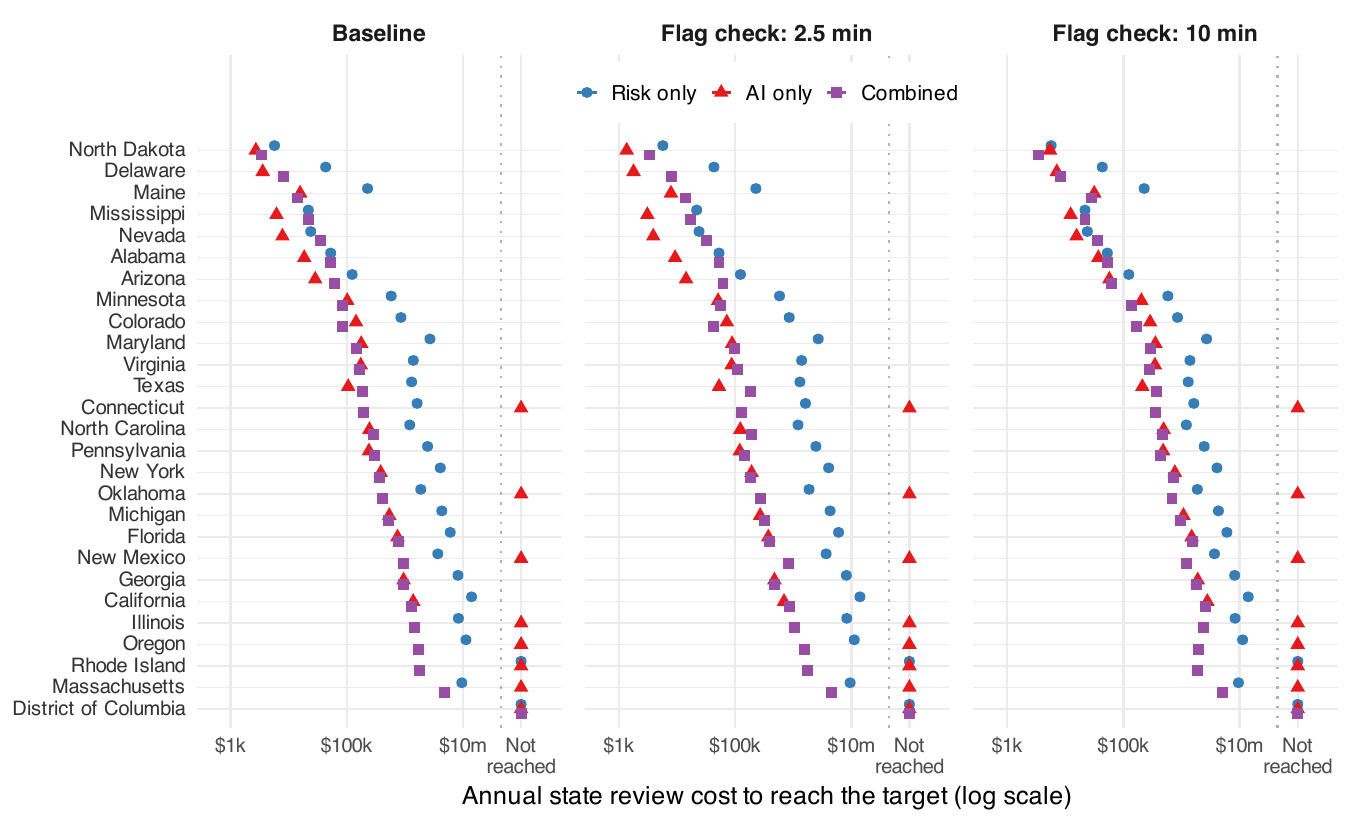}
\caption{Review costs under shorter and longer flag checks.
Risk-only review is unchanged; AI-only and combined review become more costly
as checks take longer.}
\label{fig:sensitivity-flag-time}
\end{figure}

\clearpage
\subsection{Time spent on an ordinary category check}
\label{apx:sensitivity-category-time}

We change ordinary review time per category
from \stat{review.risk.minutes} minutes
to \stat{apx.category.time.low} or \stat{apx.category.time.high} minutes,
holding correction success fixed.
Risk-only costs scale directly with the duration of a check.
Combined review is affected only when it proceeds beyond flags
to ordinary reviews of unflagged cases; AI-only costs are unchanged.
The jurisdictions reaching the target remain the same.
Among jurisdictions where both reach the target,
risk-only review costs a median of \stat{apx.ratio.risk_time_low} times as much
as combined review with shorter category checks,
and \stat{apx.ratio.risk_time_high} times as much with longer checks.

\begin{figure}[!htbp]
\centering
\includegraphics[width=\linewidth]{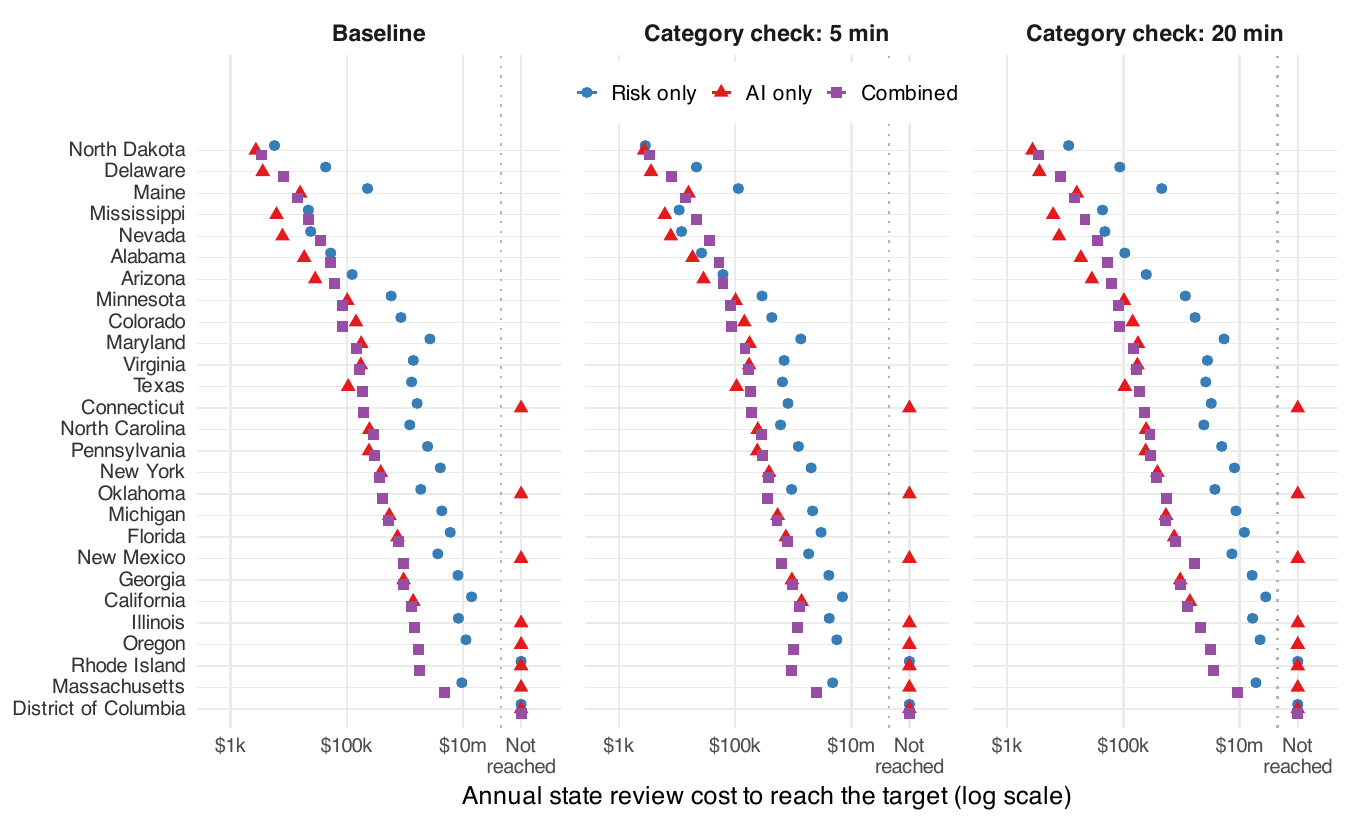}
\caption{Review costs under shorter and longer ordinary category checks.
The comparison is most favorable to risk-only review when these checks are faster.}
\label{fig:sensitivity-category-time}
\end{figure}

\clearpage
\subsection{Risk-score accuracy}
\label{apx:sensitivity-risk-scores}

We augment the original risk score $p$ with hypothetical information $U$
about whether a case contains an above-tolerance error ($Y$).
Conditional on $Y$, $U$ is independent of $p$,
with $U\mid Y=0\sim N(0,1)$ and $U\mid Y=1\sim N(d,1)$.
The updated score is
\[
 p^*=\mathrm{logit}^{-1}\!\left(
 \mathrm{logit}(p)+dU-\frac{d^2}{2}\right).
\]
We use $d=\stat{apx.score.signal.low}$ and $d=\stat{apx.score.signal.high}$,
each with \stat{apx.score.seeds} draws.
Only case rankings change; errors, correction outcomes,
and category priorities remain fixed.
These signals represent hypothetical improvements, not newly fitted models.

Among initially above-threshold jurisdictions,
the median within-jurisdiction area under the ROC curve rises
from \stat{apx.auc.baseline} to \stat{apx.auc.signal_1}
and \stat{apx.auc.signal_2}, respectively
(taking the median across draws).
Even with the stronger signal, risk-only review costs a median of
\stat{apx.ratio.strong.low}--\stat{apx.ratio.strong.high} times as much
as combined review across draws,
comparing jurisdictions where both reach the target.

\begin{figure}[!htbp]
\centering
\includegraphics[width=\linewidth]{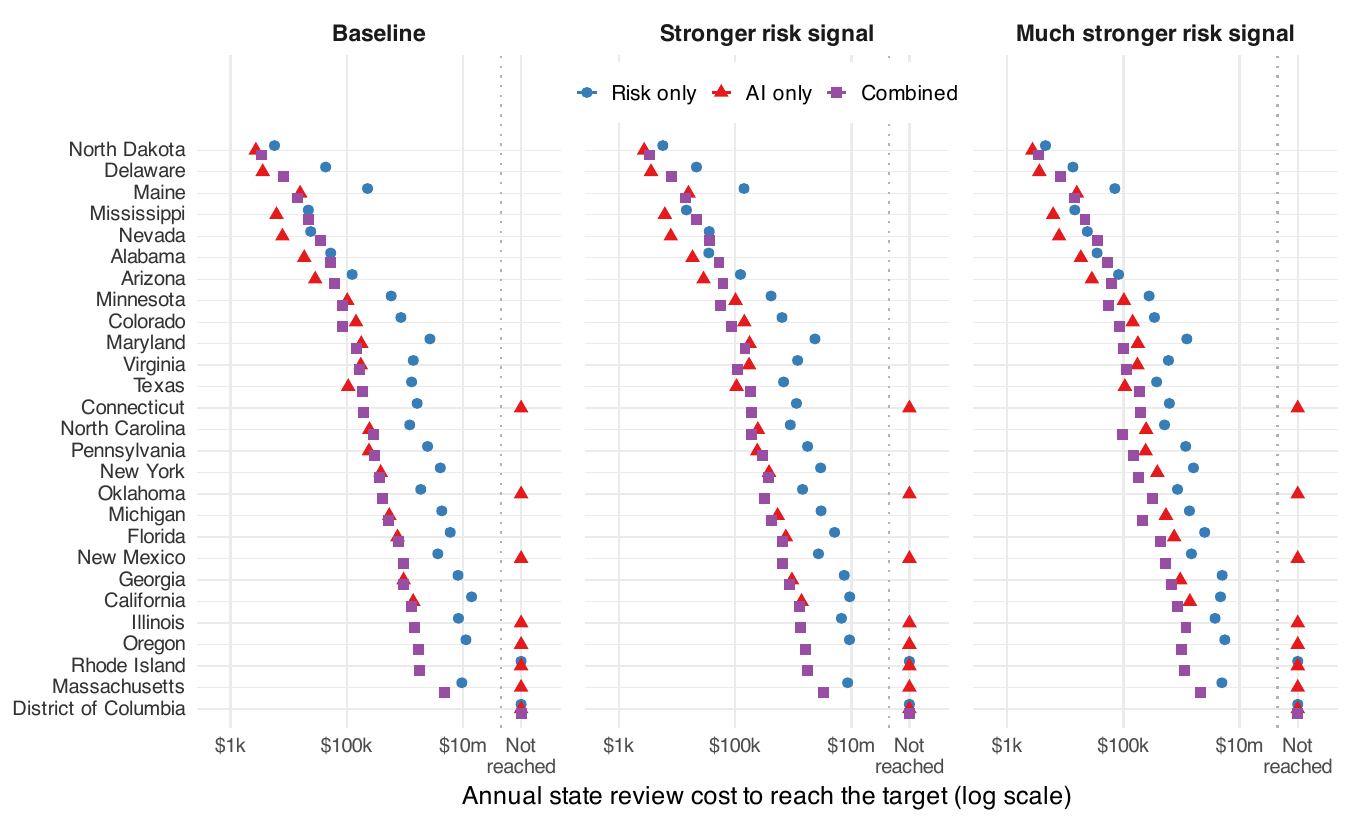}
\caption{Review costs with more informative hypothetical risk scores.
Points show median costs across \stat{apx.score.seeds} draws;
horizontal segments show their range, not confidence intervals.
AI-only review does not use risk scores and is unchanged.}
\label{fig:sensitivity-risk-signal}
\end{figure}

\clearpage
We also consider random case rankings
and rankings that reveal exactly which cases contain an above-tolerance error.
In the latter scenario, all cases with errors come first,
but the scores do not reveal the size or category of an error;
cases with the same score share the available review capacity.
Even with this knowledge, risk-only review costs a median of
\stat{apx.ratio.known_error_status} times as much as combined review
where both reach the threshold.
Changing rankings does not change the errors that can ultimately be corrected
when all reviews are completed,
so it changes costs rather than the set of jurisdictions reaching the target.

\begin{figure}[!htbp]
\centering
\includegraphics[width=\linewidth]{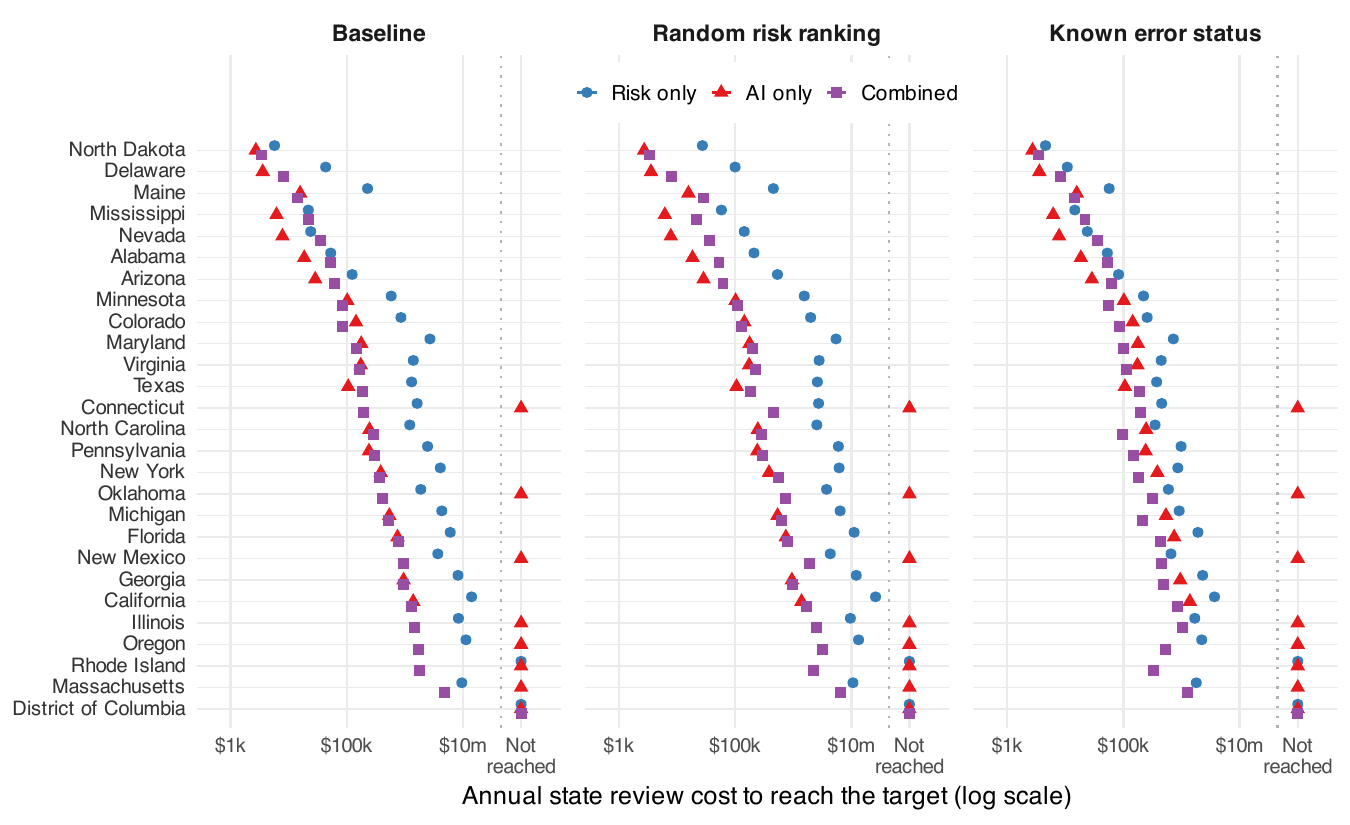}
\caption{Review costs with random case rankings or known error status.
For random rankings, points show medians across \stat{apx.score.seeds} draws
and segments show the range.
Known error status identifies cases with errors, not the errors' locations or sizes.}
\label{fig:sensitivity-risk-extremes}
\end{figure}

\clearpage
\subsection{The order of AI flag checks}
\label{apx:sensitivity-ai-order}

The main AI-only approach checks flags in categories
with the largest historical error-dollar totals for the state.
We compare this with ranking categories by historical correctable dollars
per expected hour of flag review,
using the same FY\stat{training.years} records,
and with reviewing flags in random order.
The historical-return ordering accounts for the number of flags a category
is expected to generate, rather than its total error dollars alone.
Random ordering assigns the same review fraction to flags in every category.
These variations change only standalone AI review;
the risk-only and combined approaches remain as in the main simulation.

\begin{figure}[!htbp]
\centering
\includegraphics[width=\linewidth]{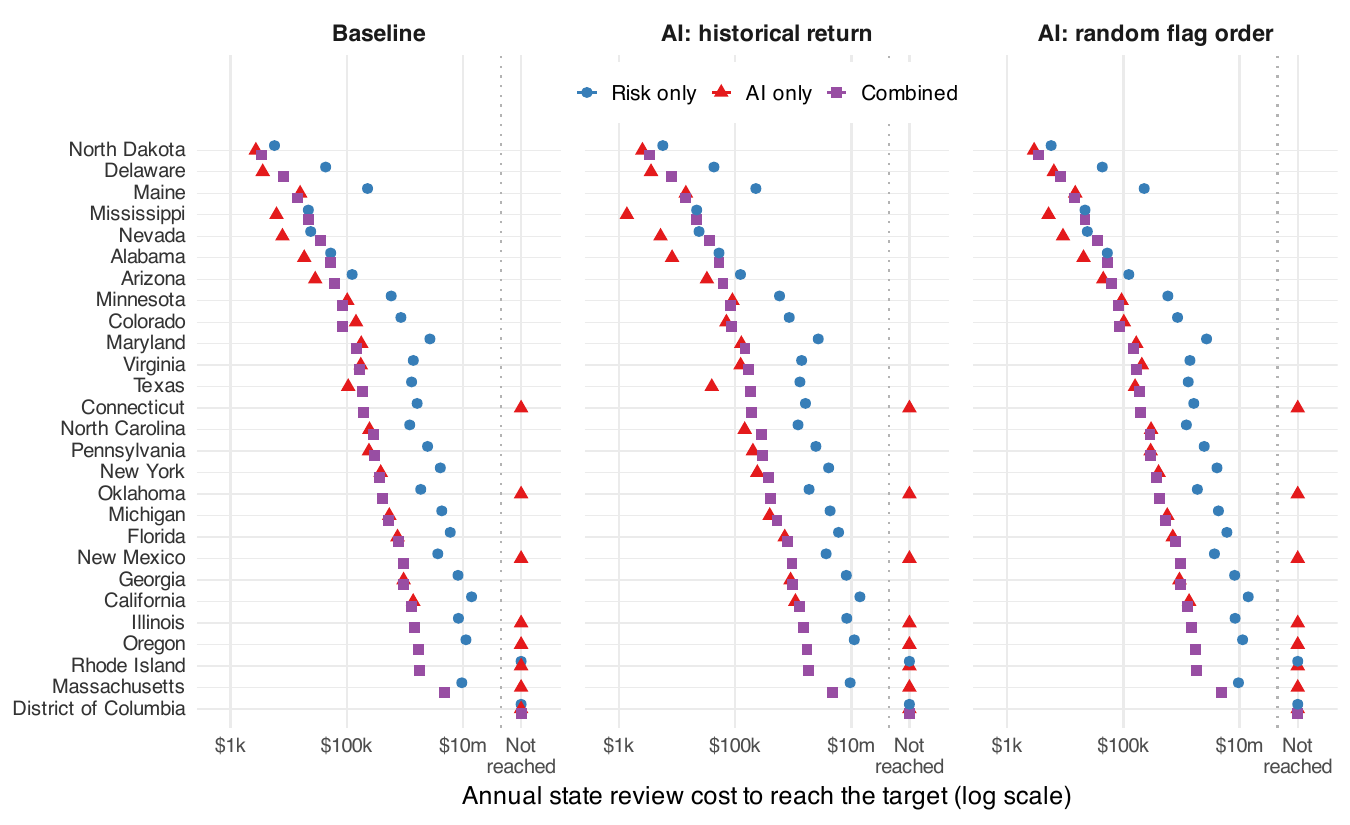}
\caption{Review costs under alternative orders for standalone AI flag checks.
``Historical return'' ranks categories by historical correctable dollars
per expected hour of review. Only the AI-only series changes.}
\label{fig:sensitivity-ai-order}
\end{figure}

All orders eventually check the same flags,
so AI-only review reaches \stat{results.ai.target.count} jurisdictions
under each ordering.
The cost to reach the threshold can nevertheless differ within a jurisdiction,
because some orders reach larger errors sooner.
Figure~\ref{fig:sensitivity-ai-order} shows those differences
without selecting the best order retrospectively for each state.

\clearpage
\subsection{The order of combined review}
\label{apx:sensitivity-combined-order}

Our main combined approach checks flags first, in case risk order,
and then performs ordinary category reviews in cases without flags.
The alternative follows case risk order throughout.
Within each selected case and category,
reviewers check flags when present and perform an ordinary category review otherwise.
Thus, this alternative also allows ordinary review of unflagged categories
in cases that have flags elsewhere;
it changes both the sequence of reviews and the work performed on those cases.

The alternative brings \stat{apx.targets.combined_order_risk.combined}
jurisdictions below the threshold,
compared with \stat{results.combined.target.count} under the main approach.
It can reach the target in the District of Columbia,
where the main approach stops short,
but typically requires more review spending to reach the target elsewhere.
The additional reach therefore comes with a cost:
reviewers spend time searching categories in high-risk cases
before checking all of the specific discrepancies identified by AI.

\begin{figure}[!htbp]
\centering
\includegraphics[width=\linewidth]{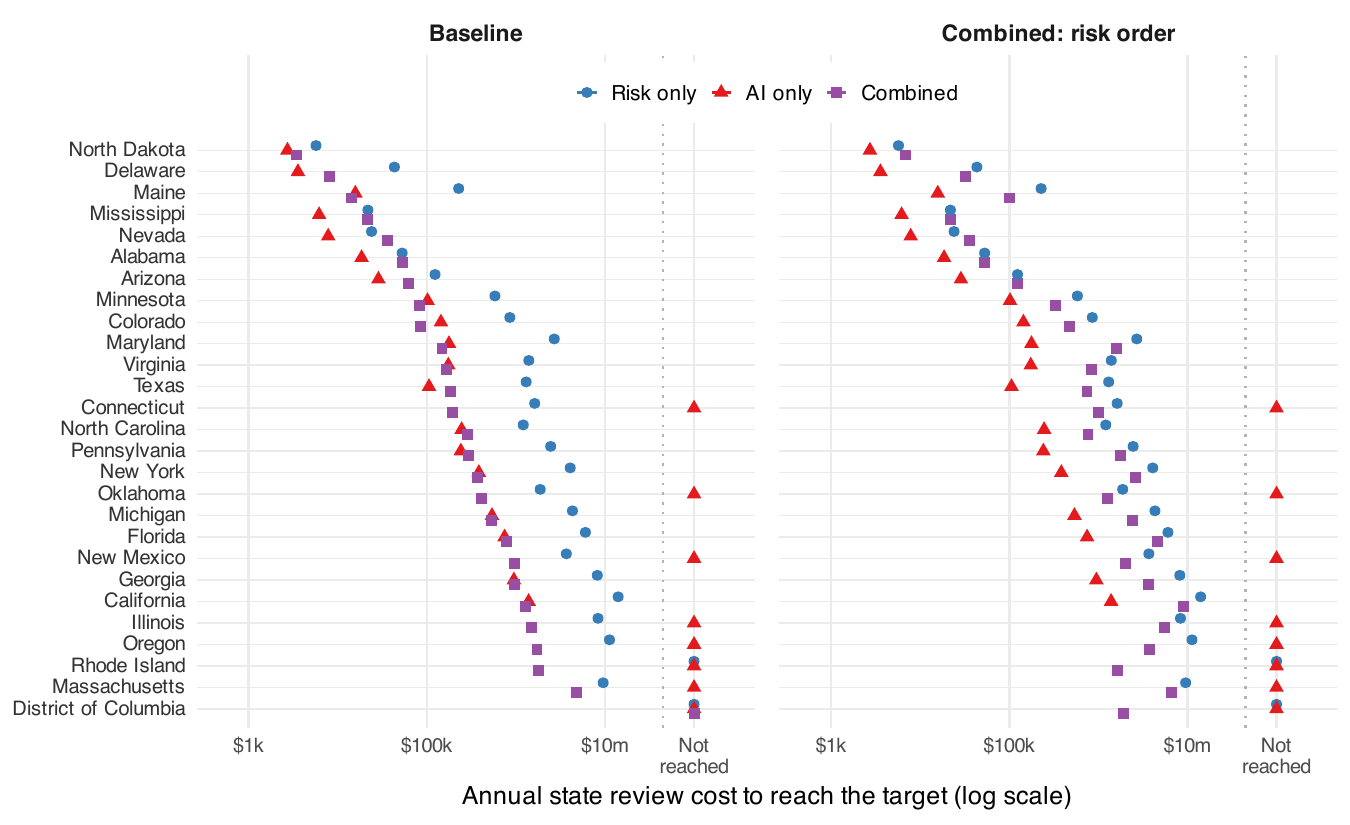}
\caption{Review costs under alternative combined-review procedures.
The main approach checks all flags before ordinary reviews of unflagged cases.
The alternative follows case risk throughout,
using ordinary review in categories without flags.}
\label{fig:sensitivity-combined-order}
\end{figure}

\end{document}

%% file: paper_stats.tex
\setstat{benefit.share.low}{5}
\setstat{results.dc.required.reduction}{57}
\setstat{results.dc.achieved.reduction}{55}
\setstat{sensitivity.scope.baseline.targets}{19}
\setstat{sensitivity.scope.broad.targets}{27}
\setstat{sensitivity.scope.narrow.targets}{14}
\setstat{sensitivity.success.low.switches}{2}
\setstat{sensitivity.success.high.switches}{4}
\setstat{sensitivity.success.baseline}{80}
\setstat{sensitivity.success.low}{60}
\setstat{sensitivity.success.high}{100}
\setstat{results.above.threshold.count}{27}
\setstat{results.combined.target.count}{26}
\setstat{results.risk.target.count}{25}
\setstat{results.ai.target.count}{19}
\setstat{results.shared.target.count}{25}
\setstat{results.combined.cheaper.count}{22}
\setstat{results.risk.combined.cost.ratio.median}{7}
\setstat{results.risk.combined.cost.ratio.range}{2--11}
\setstat{results.ca.risk.ai.time.ratio}{10}
\setstat{analysis.years}{2022--2024}
\setstat{evaluation.year}{2024}
\setstat{training.years}{2022--2023}
\setstat{per.tolerance}{56}
\setstat{risk.household.size.groups}{1, 2--3, and 4 or more}
\setstat{per.cost.sharing.threshold}{6}
\setstat{benchmark.targeting.rules.count}{three}
\setstat{benchmark.income.minutes.upper}{5}
\setstat{benchmark.expanded.minutes}{7.5}
\setstat{benchmark.projected.per.reduction}{25\%}
\setstat{benchmark.workload.reduction}{25\%}
\setstat{benchmark.errors.caught.reduction}{15\%}
\setstat{benchmark.errors.caught.retained}{85\%}
\setstat{review.risk.minutes}{10}
\setstat{review.risk.correction}{50\%}
\setstat{labor.hourly.total}{40}
\setstat{labor.hourly.state}{30}
\setstat{labor.federal.share}{25\%}
\setstat{labor.share.year}{2027}
\setstat{labor.benchmark.wage}{24.92}
\setstat{labor.benchmark.year}{2023}
\setstat{review.ai.minutes}{5}
\setstat{review.ai.correction}{80\%}
\setstat{review.ai.recall}{80\%}
\setstat{review.ai.precision}{80\%}
\setstat{review.scope.min}{1}
\setstat{review.scope.max}{12}
\setstat{review.scope.min.words}{one}
\setstat{review.scope.max.words}{twelve}
\setstat{sample.states.count}{48}
\setstat{sample.excluded.households.percent}{9.5\%}
\setstat{apx.per.sim.correlation}{0.79}
\setstat{apx.per.correlation}{0.63}
\setstat{apx.per.lower}{52}
\setstat{apx.per.count}{53}
\setstat{apx.flag.time.low}{2.5}
\setstat{apx.flag.time.high}{10}
\setstat{apx.category.time.low}{5}
\setstat{apx.category.time.high}{20}
\setstat{apx.category.success.low}{80}
\setstat{apx.category.success.high}{90}
\setstat{apx.score.seeds}{3}
\setstat{apx.score.signal.low}{1}
\setstat{apx.score.signal.high}{2}
\setstat{apx.targets.baseline.risk}{25}
\setstat{apx.targets.baseline.ai}{19}
\setstat{apx.targets.baseline.combined}{26}
\setstat{apx.targets.ai_time_low.risk}{25}
\setstat{apx.targets.ai_time_low.ai}{19}
\setstat{apx.targets.ai_time_low.combined}{26}
\setstat{apx.targets.ai_time_high.risk}{25}
\setstat{apx.targets.ai_time_high.ai}{19}
\setstat{apx.targets.ai_time_high.combined}{26}
\setstat{apx.targets.risk_time_low.risk}{25}
\setstat{apx.targets.risk_time_low.ai}{19}
\setstat{apx.targets.risk_time_low.combined}{26}
\setstat{apx.targets.risk_time_high.risk}{25}
\setstat{apx.targets.risk_time_high.ai}{19}
\setstat{apx.targets.risk_time_high.combined}{26}
\setstat{apx.targets.ai_success_low.risk}{25}
\setstat{apx.targets.ai_success_low.ai}{17}
\setstat{apx.targets.ai_success_low.combined}{24}
\setstat{apx.targets.ai_success_high.risk}{25}
\setstat{apx.targets.ai_success_high.ai}{21}
\setstat{apx.targets.ai_success_high.combined}{27}
\setstat{apx.targets.risk_success_high.risk}{27}
\setstat{apx.targets.risk_success_high.ai}{19}
\setstat{apx.targets.risk_success_high.combined}{27}
\setstat{apx.targets.risk_success_90.risk}{27}
\setstat{apx.targets.risk_success_90.ai}{19}
\setstat{apx.targets.risk_success_90.combined}{27}
\setstat{apx.targets.ai_recall_low.risk}{25}
\setstat{apx.targets.ai_recall_low.ai}{17}
\setstat{apx.targets.ai_recall_low.combined}{25}
\setstat{apx.targets.ai_recall_high.risk}{25}
\setstat{apx.targets.ai_recall_high.ai}{21}
\setstat{apx.targets.ai_recall_high.combined}{27}
\setstat{apx.targets.ai_scope_narrow.risk}{25}
\setstat{apx.targets.ai_scope_narrow.ai}{14}
\setstat{apx.targets.ai_scope_narrow.combined}{25}
\setstat{apx.targets.ai_scope_broad.risk}{25}
\setstat{apx.targets.ai_scope_broad.ai}{27}
\setstat{apx.targets.ai_scope_broad.combined}{27}
\setstat{apx.targets.random.risk}{25}
\setstat{apx.targets.random.ai}{19}
\setstat{apx.targets.random.combined}{26}
\setstat{apx.targets.signal_1.risk}{25}
\setstat{apx.targets.signal_1.ai}{19}
\setstat{apx.targets.signal_1.combined}{26}
\setstat{apx.targets.signal_2.risk}{25}
\setstat{apx.targets.signal_2.ai}{19}
\setstat{apx.targets.signal_2.combined}{26}
\setstat{apx.targets.known_error_status.risk}{25}
\setstat{apx.targets.known_error_status.ai}{19}
\setstat{apx.targets.known_error_status.combined}{26}
\setstat{apx.targets.ai_order_return.ai}{19}
\setstat{apx.targets.ai_order_random.ai}{19}
\setstat{apx.targets.ai_order_random.combined}{26}
\setstat{apx.targets.ai_order_return.combined}{26}
\setstat{apx.targets.ai_order_random.risk}{25}
\setstat{apx.targets.ai_order_return.risk}{25}
\setstat{apx.targets.combined_order_risk.combined}{27}
\setstat{apx.targets.combined_order_risk.risk}{25}
\setstat{apx.targets.combined_order_risk.ai}{19}
\setstat{apx.ratio.baseline}{7.0}
\setstat{apx.ratio.ai_time_low}{8.0}
\setstat{apx.ratio.ai_time_high}{4.2}
\setstat{apx.ratio.risk_time_low}{3.5}
\setstat{apx.ratio.risk_time_high}{14.0}
\setstat{apx.ratio.ai_success_low}{4.6}
\setstat{apx.ratio.ai_success_high}{8.3}
\setstat{apx.ratio.risk_success_high}{2.8}
\setstat{apx.ratio.risk_success_90}{2.5}
\setstat{apx.ratio.ai_recall_low}{5.3}
\setstat{apx.ratio.ai_recall_high}{7.0}
\setstat{apx.ratio.ai_scope_narrow}{2.7}
\setstat{apx.ratio.ai_scope_broad}{11.0}
\setstat{apx.ratio.random}{10.2}
\setstat{apx.ratio.signal_1}{5.8}
\setstat{apx.ratio.signal_2}{4.0}
\setstat{apx.ratio.known_error_status}{3.0}
\setstat{apx.ratio.ai_order_random}{7.0}
\setstat{apx.ratio.ai_order_return}{7.0}
\setstat{apx.ratio.combined_order_risk}{1.6}
\setstat{apx.auc.baseline}{0.77}
\setstat{apx.auc.random}{0.50}
\setstat{apx.auc.signal_1}{0.85}
\setstat{apx.auc.signal_2}{0.95}
\setstat{apx.auc.known_error_status}{1.00}
\setstat{apx.ratio.strong.low}{4.0}
\setstat{apx.ratio.strong.high}{4.7}
\setstat{results.risk.combined.cost.ratio.median.words}{seven}
\setstat{simulation.combined.replicates}{200}

%% file: ai_coverage_table.tex
\begin{tabular}{@{}p{0.49\linewidth}p{0.47\linewidth}@{}}
\toprule
\textbf{Assumed detectable} & \textbf{Assumed not detectable} \\
\midrule
\textbf{2}: Incomplete or incorrect information provided; agency not required to verify.\par\smallskip
\textbf{10}: Policy incorrectly applied.\par\smallskip
\textbf{12}: Reported information disregarded or not applied.\par\smallskip
\textbf{14}: Agency failed to follow up on inconsistent or incomplete information.\par\smallskip
\textbf{15}: Agency failed to follow up on impending changes.\par\smallskip
\textbf{16}: Agency failed to verify required information.\par\smallskip
\textbf{17}: Computer programming error.\par\smallskip
\textbf{18}: Data entry and/or coding error.\par\smallskip
\textbf{19}: Mass change (error due to problem with computer-generated mass change).\par\smallskip
\textbf{20}: Arithmetic computation error.\par\smallskip
\textbf{21}: Computer user error.\par\smallskip
\textbf{22}: Agency budgeted an incorrect amount.\par\smallskip
\textbf{23}: Agency failed to follow recertification procedure related to notices/forms.\par\smallskip
\textbf{24}: Agency failed to follow recertification procedure related to interviews.\par\smallskip
\textbf{25}: Agency failed to follow recertification procedure related to time frames.\par\smallskip\textbf{Missing cause}: Assumed detectable. & \textbf{1}: Information not reported.\par\smallskip
\textbf{3}: Information withheld by client (case referred for Intentional Program Violation [IPV] investigation).\par\smallskip
\textbf{4}: Incorrect information provided by client (case referred for IPV investigation).\par\smallskip
\textbf{7}: Inaccurate information reported by collateral contact.\par\smallskip
\textbf{8}: Acted on incorrect federal computer match information not requiring verification (such variance is excluded from error determination but must be recorded).\par\smallskip
\textbf{26}: Change was not required to be reported by the client or acted upon by the State agency based on time frames and reporting requirements.\par\smallskip
\textbf{99}: Other. \\
\bottomrule
\end{tabular}

%% file: review_categories_table.tex
\textbf{Earned income} & \texttt{311}: Wages and salaries; \texttt{314}: Other earned income; \texttt{321}: Earned income deductions. \\ \addlinespace[5pt]
\textbf{Unearned income} & \texttt{331}: RSDI benefits; \texttt{332}: Veterans' benefits; \texttt{333}: SSI and/or State SSI supplement; \texttt{334}: Unemployment compensation; \texttt{335}: Workers' compensation; \texttt{336}: Other government benefits; \texttt{342}: Contributions; \texttt{343}: Deemed income; \texttt{344}: TANF, PA, or GA; \texttt{345}: Educational grants/scholarships/loans; \texttt{346}: Other unearned income; \texttt{350}: Child support payments received from absent parent. \\ \addlinespace[5pt]
\textbf{Shelter and utilities} & \texttt{362}: Homeless shelter deduction; \texttt{363}: Shelter deduction; \texttt{364}: Standard utility allowance. \\ \addlinespace[5pt]
\textbf{Household composition} & \texttt{140}: Residency; \texttt{150}: Unit composition; \texttt{170}: Social Security number. \\ \addlinespace[5pt]
\textbf{Self-employment} & \texttt{312}: Self-employment. \\ \addlinespace[5pt]
\textbf{Medical expenses} & \texttt{365}: Medical expense deductions. \\ \addlinespace[5pt]
\textbf{Care and child support} & \texttt{323}: Dependent care deduction; \texttt{366}: Child support payment deduction. \\ \addlinespace[5pt]
\textbf{Citizenship and students} & \texttt{111}: Student status; \texttt{130}: Citizenship and noncitizen status. \\ \addlinespace[5pt]
\textbf{Other computation and reporting} & \texttt{520}: Arithmetic computation; \texttt{530}: Transitional benefits; \texttt{540}: Missing reports; \texttt{542}: Expired certification period; \texttt{560}: Reporting systems; \texttt{810}: SNAP simplification project; \texttt{820}: Demonstration projects. \\ \addlinespace[5pt]
\textbf{Work and disqualifications} & \texttt{151}: Recipient disqualification; \texttt{160}: Employment and training programs; \texttt{161}: Time-limited participation; \texttt{162}: Work registration requirements; \texttt{163}: Voluntary quit/reduced work effort; \texttt{164}: Workfare and comparable workfare; \texttt{165}: Employment status/job availability; \texttt{166}: Acceptance of employment. \\ \addlinespace[5pt]
\textbf{Income totals and standard deduction} & \texttt{361}: Standard deduction; \texttt{371}: Combined gross income; \texttt{372}: Combined net income. \\ \addlinespace[5pt]
\textbf{Assets} & \texttt{211}: Bank accounts or cash on hand; \texttt{212}: Nonrecurring lump-sum payment (includes non-substantial lottery or gambling winnings); \texttt{213}: Other liquid assets; \texttt{214}: Substantial lottery or gambling winnings; \texttt{221}: Real property; \texttt{222}: Vehicles; \texttt{224}: Other nonliquid resources; \texttt{225}: Combined resources. \\ \addlinespace[5pt]